\documentclass[prd,showpacs,preprintnumbers,amsmath,amssymb,nofootinbib]{revtex4}

\usepackage{graphicx}
\usepackage{dcolumn}
\usepackage{bm}

\newcommand{\Od}{{\cal O}}

\newcommand{\ematriz}[3]{\left\langle {#1} \left|{#2}\right|{#3}\right\rangle}

\newcommand{\parent}[1]{\left({#1}\right)}

\newcommand{\condtwo}{\langle \bar q q \rangle}
\newcommand{\condtwol}{\langle (\bar q q)_l \rangle}
\newcommand{\condfour}{\langle (\bar q q)^2 \rangle}
\newcommand{\condn}{\langle (\bar q q)^n \rangle}
\newcommand{\quarkcor}{\langle {\cal T} (\bar q q)(x) (\bar q
q)(0)\rangle}

\newcommand{\condtwoT}{\langle \bar q q \rangle_{T}}
\newcommand{\condsT}{\langle \bar s s \rangle_{T}}
\newcommand{\condtwolT}{\langle (\bar q q)_l \rangle_{T}}
\newcommand{\condfourT}{\langle (\bar q q)^2 \rangle_{T}}

\newcommand{\quarkcorT}{\langle {\cal T} (\bar q q)(x) (\bar q
q)(0)\rangle_{T}}
\newcommand{\quarkcorlT}{\langle {\cal T} (\bar q q)_l(x) (\bar q
q)_l(0)\rangle_{T}}

\newcommand{\be}{\begin{equation}}
\newcommand{\ee}{\end{equation}}
\newcommand{\ba}{\begin{eqnarray}}
\newcommand{\ea}{\end{eqnarray}}

\newcommand{\gsim}{\raise.3ex\hbox{$>$\kern-.75em\lower1ex\hbox{$\sim$}}}
\newcommand{\lsim}{\raise.3ex\hbox{$<$\kern-.75em\lower1ex\hbox{$\sim$}}}

\begin{document}

\title{Scalar susceptibilities and four-quark condensates in the meson gas within Chiral Perturbation Theory}


\author{A. G\'omez Nicola}
\email{gomez@fis.ucm.es} \affiliation{Departamento de F\'{\i}sica
Te\'orica II. Universidad Complutense. 28040 Madrid. Spain.}
\author{J.R. Pel\'aez}
\email{jrpelaez@fis.ucm.es} \affiliation{Departamento de
F\'{\i}sica Te\'orica II. Universidad Complutense. 28040 Madrid. Spain.}
\author{J. Ruiz de Elvira}\email{jacobore@rect.ucm.es}
\affiliation{Departamento de F\'{\i}sica Te\'orica II. Universidad
Complutense. 28040 Madrid. Spain.}

\begin{abstract}
We analyze the properties of four-quark condensates and scalar susceptibilities in the meson gas, within finite temperature Chiral Perturbation Theory (ChPT). The breaking of the factorization hypothesis does not allow for a finite four-quark condensate and its use as an order parameter, except in the chiral limit. This is rigorously obtained within ChPT and is therefore a model-independent result. Factorization only holds formally in the large $N_c$ limit and breaks up at finite temperature even in the chiral limit. Nevertheless, the factorization breaking terms are precisely those needed to yield a finite scalar susceptibility, deeply connected to chiral symmetry restoration. Actually, we provide the full result for the $SU(3)$ quark condensate to NNLO in ChPT, thus extending previous results to include kaon and eta interactions. This allows to check the effect of those corrections compared to previous approaches and the uncertainties due to low-energy constants.
We provide a detailed analysis of scalar susceptibilities in the $SU(3)$ meson gas, including a comparison between the pure ChPT approach and the virial expansion, where the unitarization of pion scattering is crucial to achieve a more reliable prediction.
Through the analysis of the interactions within this approach, we have found that the role of the $\sigma$ resonance
is largely canceled with the scalar isospin two channel interaction, leaving the $\rho(770)$ as the main contribution.
 Special attention is paid to
the evolution towards chiral restoration, as well as to the comparison with recent lattice analysis.
\end{abstract}



\maketitle

\section{Introduction}

Chiral symmetry restoration \cite{PisWil84} is a very relevant ingredient in our present understanding of hadronic physics under extreme conditions of temperature and density and has been one of the main motivations for  the development of the heavy-ion and nuclear matter experimental programs, which are still producing new results in facilities such as RHIC, CERN (ALICE) and FAIR. In parallel, the improvement of lattice data at finite temperature, performed by different groups \cite{Bernard:2004je,Aoki:2006br,Aoki:2006we,Cheng:2009zi,Aoki:2009sc,Ejiri:2009ac,Borsanyi:2010bp},
  has contributed considerably to clarify the main properties of the chiral symmetry transition, which is believed to take place in the same range as the deconfinement one. Nowadays, there is a fair consistency between  lattice simulations performed with different methods, pointing towards  a crossover-like transition for $N_f=3$ (2+1 flavors in the physical case), becoming  of second order  for $N_f=2$ (in the $O(4)$ universality class) and first order in the degenerate case of three equal flavors. This behavior corresponds to vanishing baryon chemical potential and the transition temperature lies within the range $T_c\sim$ 150-175 MeV. It is important to remark that in the physical 2+1 case analyzed on the lattice, the transition being a crossover means that one should really talk about a transition range rather than a critical temperature, and that range can be established by looking at different order-like parameters, which can give different values for $T_c$. The  chief parameters used in lattice analysis are the quark condensate and susceptibilities, defined as first derivatives of the quark condensates with respect to quark masses. Susceptibilities   measure fluctuations of the associated order parameter and can be expressed in terms of current correlators. Thus, the scalar susceptibility, related to the quark condensate, is expected to grow faster just below the transition.

It is important to provide an accurate analytical description of the physics below the chiral transition,  to compare  with experimental data and  to confront the lattice results in the continuum. On the one hand, a particularly useful approach has been the Hadron Resonance Gas (HRG), which has proven to be quite successful to describe thermodynamic quantities, i.e., those that can be derived directly from the free energy density, as compared to lattice data \cite{Karsch:03,Huovinen:2009yb,Borsanyi:2010bp}. Within the usual HRG approach,  the free contribution of  all known physical hadron states to the partition function is considered. Models including resonance widths  and hadron interactions  improve the  HRG  approach description of hadron production experimental data \cite{Andronic:2005yp,Andronic:2008gu} and lattice results \cite{Andronic:2012ut}.

On the other hand, Chiral Perturbation Theory (ChPT) \cite{we79,Gasser:1983yg,Gasser:1984gg} allows to describe the low-temperature meson gas \cite{Gasser:1986vb,Gerber:1988tt} in a model-independent and systematic way, for $SU(N_f)_L\times SU(N_f)_R\rightarrow SU(N_f)_V$ symmetry breaking with $N_f=2,3$ light flavors. In particular, interactions can be included in the most general way compatible both with the underlying symmetries and with meson-meson  scattering data.  Although ChPT includes only the lightest degrees of freedom ($\pi, K, \eta$) and hence is limited to low and moderate temperatures, it provides model-independent results. In fact, in early studies of the partition function \cite{Gerber:1988tt}, extrapolations of the condensate and other quantities to the transition region give a qualitatively reasonable description of the relevant physics. In addition, there have been some relevant developments on the use of ChPT results
in the description of the meson gas properties. For instance, it is known that
the Inverse Amplitude Method \cite{Truong:1988zp,Dobado:1992ha,GomezNicola:2007qj},
which is a dispersive method to unitarize ChPT without introducing spurious parameters,
allows the generation of light meson resonances ( the $\sigma$ or $f_0(500)$, the $\rho(770)$, ...). Within that unitarized ChPT scheme,  useful results have been developed for the temperature and density dependence of the lightest resonances and their connection to chiral restoration and chiral partner degeneration \cite{Dobado:2002xf,FernandezFraile:2007fv,Cabrera:2008tja} as well as a phenomenologically successful description of transport coefficients \cite{FernandezFraile:2009mi}. It is particularly relevant to recall the virial expansion \cite{Dashen,Gerber:1988tt,Welke:1990za,Venugopalan:1992hy,Nyffeler:1993iz,Dobado:1998tv,Pelaez:2002xf} which allows to parametrize efficiently the effect of meson interactions in the partition function within a dilute gas description, valid below the transition. States of more energy are weighted by Boltzmann factors and then become more relevant as the system approaches the transition. Thus, in this approach, it is more important to include accurately the interaction of the lightest mesons, e.g. via unitarization, while the heavier ones can be added as free states. Actually, the HRG approach with just free states is nothing but the leading order in the virial expansion.
An alternative  approach, also within unitarized ChPT, is  to take into account the temperature or density dependence of the phase shifts, which would make the $\sigma$ resonance pole move towards the real axis precisely as a signature of chiral restoration \cite{FernandezFraile:2007fv,Cabrera:2008tja}.

In this work we will explore some additional properties of the meson gas within the ChPT framework, regarding in particular quark condensates and susceptibilities. Our study will be deeply connected to the  analysis of four-quark correlators of the type $\quarkcor$ and the factorization hypothesis, thus extending our previous work at zero temperature \cite{GomezNicola:2010tb}. This hypothesis states that four-quark condensates factorize into two-quark condensates squared $\condfour\sim \condtwo^2$ with the same quantum numbers. We have shown in \cite{GomezNicola:2010tb} that this hypothesis fails to next to next to leading order (NNLO) in ChPT. Here, we will show that the same holds at finite temperature, preventing the use of four-quark condensates as order parameters since the factorization breaking terms diverge. In the derivation we will in turn obtain the $SU(3)$ thermal quark condensate in ChPT including all the relevant meson interactions, which extends previous calculations of the condensate to this order which considered free kaons and etas \cite{Gerber:1988tt}. The non-factorizing scalar four-quark correlator gives rise to the scalar susceptibility, allowing for a direct check of the calculation. Once the connection with factorization is established, we will perform a detailed analysis of the scalar susceptibilities in the virial approach, with and without including unitarized amplitudes, extending previous works in the literature and serving as a test of the robustness of the ChPT results, which together with their model independence makes them a useful prediction for low and moderate temperatures below the transition. In this respect, it is particularly relevant to note that unitarized ChPT is able to provide a relatively good description of the quark mass dependence of the resonant states obtained through
unitarization \cite{qmassUChPT,Nebreda:2010wv,Nebreda:2011di}, particularly robust in the case of the lightest scalar.

The paper is organized as follows. After fixing our notation and definitions (section \ref{sec:gendef}), in section \ref{sec:fourq} we present our calculation of the relevant four-quark correlators for two and three flavors at finite temperature. The details  are given for $N_f=3$.   The factorization hypothesis is then examined in section \ref{sec:factorization}, where we also comment on the large-$N_c$ limit. In section \ref{sec:sus} we establish the connection with the scalar susceptibility.  Different effects in the
 ChPT condensates and susceptibilities are discussed in section \ref{sec:pheno}, while section \ref{sec:vir} is devoted to the analysis
  within the virial approach, to compare with previous analysis and to study the role of interactions. We pay special attention to the comparison with
   lattice data (section \ref{sec:latcom})  and to study its behavior as the system approaches  chiral restoration.
In Appendix \ref{sec:su2res} we collect some of the $SU(2)$ results, while in Appendix \ref{app:renquark} we provide the detailed expressions for the thermal quark condensates to NNLO in ChPT.

\section{QCD Condensates, Susceptibilities and four-quark correlators in Chiral Perturbation Theory}

\subsection{General definitions}
\label{sec:gendef}

Let us start from the QCD Euclidean Lagrangian including scalar sources:

\begin{equation}
{\cal L}_{QCD}[q,\bar q, s(x)]=\bar q\parent{i \not \! \! D-s(x)}q+\cdots,
 \end{equation}
where the rest of the Lagrangian indicated by dots is irrelevant for our
purposes, the $(-,-,-,-)$ metric is used and a sum over $N_f$ flavor, $N_c$
colors and Dirac indices is implicit.

The physical QCD Lagrangian
corresponds to setting $s(x)={\cal M}$, the quark
mass matrix. In the three flavor case
 $\mathcal M$=diag($m_u$,$m_d$,$m_s$), where $m_u$, $m_d$ and $m_s$ correspond to the up, down and
strange mass respectively. For simplicity, we will work in the isospin limit,
so that $m_u=m_d=m$ and $s(x)=$diag$(s_0(x),s_0(x),s_s(x)$).

We will follow the external source method \cite{Gasser:1984gg} to deal with the different two-quark ($\bar q q$) and four-quark ($\bar q q \bar q q$) correlators of interest. Consider first the quark condensates at finite temperature:

\begin{eqnarray}
\condtwo_T \equiv \langle \bar u u + \bar d d +\bar s s\rangle_T\equiv\frac{1}{Z_{QCD}[{\cal M}]}\int{{\cal D} \bar q {\cal D} q}\cdots\,\bar q q \,\exp{\int_E d^4x {\cal L}_{QCD}[\bar q,q,s(x),\cdots]},
\end{eqnarray}
where   $Z_{\mathrm{QCD}}[\mathcal M]$ is the partition function.
 In the above equation  $\int_E{d^4x}=\int_0^\beta d\tau\int d^3\vec{x}$ is the Euclidean (imaginary-time $t=-i \tau$) version of the Minkowski volume $i\int{d^4x}$, the averaging is performed over the thermal ensemble (an asymmetric box with imaginary time extension of $\beta=1/T \ll L$ and $V=L^3$). The dots indicate the dependence on the rest of the QCD Lagrangian fields, not relevant for our purposes. Similar equations hold for the light condensate $\condtwolT\equiv\langle \bar u u + \bar d d \rangle_T$. Recall that the light sector ($u,d$) is the most relevant one concerning chiral symmetry restoration, mostly  due to the heavier strange mass.

We consider the effective low-temperature representation given by Chiral
Perturbation Theory (ChPT) \cite{we79,Gasser:1983yg,Gasser:1984gg} of the QCD generating functional, built from chiral symmetry invariance as an expansion in external momenta and quark masses:
\begin{eqnarray}
Z_{QCD}[s]&\simeq&Z_{eff}[s]=\int {\cal D}\phi^a \exp{\int_E d^4 x {\cal L}_{eff}[\phi^a,s(x)]},
\nonumber\\
{\cal L}_{eff}&=&{\cal L}_2+{\cal L}_4+{\cal L}_6\ldots,
\label{zeff}
\end{eqnarray}

We thus have:

\begin{equation}\label{cond Z}
\condtwo_T = \frac{-1}{Z_{QCD}[{\cal M}]}\left(\frac{\delta }{\delta s_0(x)}+\frac{\delta} {\delta s_s(x)}\right)Z_{QCD}[s]\Bigg\vert_{s={\cal M}}\simeq \frac{-1}{Z_{eff}[{\cal M}]}\left(\frac{\delta }{\delta s_0(x)}+\frac{\delta }{\delta s_s(x)}\right)Z_{eff}[s]\Bigg\vert_{s={\cal M}},
\end{equation}
where ${\cal L}_{eff}$  is the most general one made out of pion, kaon and eta fields $\phi^a$, that
respects the QCD chiral symmetry breaking pattern. These particles are the QCD
low-energy degrees of freedom since they are Nambu-Goldstone bosons (NGB) of the QCD
spontaneous chiral symmetry breaking. The subscript in the effective Lagrangian indicates the order in the
ChPT derivative and mass expansion ${\cal L}={\cal O} (p^{2k})$ over a typical scale $\Lambda_\chi \sim$ 1 GeV, and $\phi^a$ denote
the NGB fields. Since
the $u, d, s$ quark masses are  small compared with $\Lambda_\chi$, they are introduced
as perturbations, giving rise to the $\pi$, $K$ and $\eta$ masses, counted as $O(p^2)$. At each
order, ${\cal L}_{eff}$ is the sum of all terms compatible with the symmetries, multiplied
by chiral parameters, which absorb loop divergences order by order, yielding finite
results. ChPT is thus the quantum effective field theory of QCD, and it allows for
a systematic and model independent analysis of low-energy mesonic processes.
The NGB fields are usually collected in the $SU(N_f)$ matrix
$U=\exp[i\lambda_a\phi^a/F]$, where, in the
$N_f=3$ case, $\lambda_a$ are the Gell-Mann matrices.
The Lagrangian $\mathcal L_2$ is the non-linear sigma model:
\begin{equation}
\mathcal L_2=\frac{F^2}{4}\mathrm{Tr}\left[\partial_\mu U^\dagger\partial^\mu U +\chi(U+U^\dagger)\right],
\end{equation}
with $\chi=2B_0 s(x)$, while $F$ is the pion decay constant in the chiral limit. When $s(x)=\mathcal M$, the following lowest order SU(3) relations hold: $\condtwo=-3B_0 F^2$, $M_{0\pi}^2=2B_0 m$, $M_{0K}^2=B_0 (m+m_s)$ and
$M_{0\eta}^2=\frac{2}{3} B_0(m+2m_s$). The ChPT power counting can be formally traced in terms of the counting in $1/F^2$ and so we will do in the following. The Lagrangians $\mathcal L_4$ and $\mathcal L_6$ are given in
\cite{Gasser:1984gg} and \cite{Bijnens:1999sh} respectively, and contain the
so-called low energy constants (LEC), $L_i$ and $H_i$ (the latter are contact terms without NGB fields) for $\mathcal L_4$ and
$C_i$ for $\mathcal L_6$.

The quark condensates (light and strange) can also be defined in terms of the free energy density $z$, which in the thermodynamic limit is given by:
\begin{equation}\label{z}
z=-\lim_{V\rightarrow \infty}{\frac{1}{\beta V}\log Z},
\end{equation}
so that:
\begin{eqnarray}\label{cond z}
\condtwolT = \frac{\partial z}{\partial m},\qquad \condsT=\frac{\partial z}{\partial m_s}, \qquad
\condtwo_T =\condtwolT+\condsT.
\end{eqnarray}

We turn now to the susceptibilities and their relation to four-quark correlators, which will play an important role in this work. Susceptibilities are defined as variations of the condensates with respect to the quark masses and are directly related to the thermal averages of four-quark operators measuring condensate fluctuations. Thus, the  euclidean light scalar (or chiral) susceptibility is given by:

\begin{eqnarray}
\chi_l (T)&=&-\frac{\partial}{\partial m} \condtwolT=-\frac{\partial^2}{\partial m^2}z=\frac{1}{\beta V}\left[\frac{1}{Z[\mathcal{M}]}\frac{\partial^2}{\partial m^2}Z[\mathcal{M}]-\parent{\frac{1}{Z[\mathcal{M}]}\frac{\partial }{\partial m}Z[\mathcal{M}]}^2\right]\nonumber \\
&=&\int_E{d^4x \left[\quarkcorlT-\condtwolT^2\right]},
\label{chil4q}\end{eqnarray}
which relates the light susceptibility (the most relevant one regarding chiral restoration) with the  four-quark correlator of the light combination $(\bar q q)_l= \bar u u + \bar d d$:

\begin{equation}\label{fourquarkcorlight}
\quarkcorlT= \frac{1}{Z[{\cal M}]}\frac{\delta}{\delta s_0(x)}\frac{\delta}{\delta s_0(0)} Z[s]\bigg\vert_{s={\cal M}}\simeq \frac{1}{Z_{eff}[{\cal M}]}\frac{\delta}{\delta s_0(x)}\frac{\delta}{\delta s_0(0)} Z_{eff}[s]\bigg\vert_{s={\cal M}},
\end{equation}
where ${\cal T}$ denotes euclidean time ordering, and so on for the strange, light-strange and full $SU(3)$ susceptibilities:

\begin{eqnarray}
\chi_s(T)&=&-\frac{\partial}{\partial m_s} \condsT=-\frac{\partial^2 z}{\partial^2 m_s}=\int_E{d^4x \left[\langle {\cal T} (\bar s s)(x) (\bar s
s)(0)\rangle_T-\condsT^2\right]}\label{chis4q},\\
\chi_{ls}(T)&=&-\frac{\partial}{\partial m_s}\condtwolT=-\frac{\partial}{\partial m} \condsT=-\frac{\partial^2 z}{\partial m \partial m_s}=\int_E{d^4x \left[\langle {\cal T}(\bar q q
)_l (x)(\bar s s) (0)\rangle_T-\condtwolT\condsT\right]}\label{chils4q},\\
\chi (T)&=&\chi_l(T)+2\chi_{ls}(T)+\chi_s(T)=\int_E{d^4x \left[\quarkcor_T-\condtwo_T^2\right]}.
\label{chitot4q}\end{eqnarray}

Note that, since the low-$T$ representation of the free energy density is finite and independent of the low-energy renormalization scale $\mu$ \cite{Gerber:1988tt}  so are the quark condensates and susceptibilities, which can be expressed as mass derivatives of $z$.  However, that is not the case of the four-quark condensates, which we define, for the different combinations of quark flavors, as:

\begin{equation}
\langle (\bar q q)_\alpha (\bar q q)_\beta \rangle_T=\lim_{x\rightarrow 0}\langle {\cal T} (\bar q q)_\alpha (x)(\bar q q)_\beta (0)\rangle_T,
\label{cond4qdef}
\end{equation}
where $(\bar q q)_\alpha$ stands for either $(\bar q q)_l$ or $(\bar s s)$.
Actually, at $T=0$ the low-energy representations of  the four-quark condensates are divergent and scale-dependent, which is consistent with Renormalization Group (RG) analyses and holds also for other definitions of the four-quark condensates from the four-quark correlators, different from that in Eq.~(\ref{cond4qdef}), as we showed in \cite{GomezNicola:2010tb}.

Our previous discussion will allow us to relate, in the finite-temperature case, susceptibilities to four-quark correlators,  which in the next section will be calculated within ChPT.  Susceptibilities can be calculated either directly as mass derivatives of the two-quark condensates, or from the four-quark correlators, as in Eq.~(\ref{chil4q}), always yielding a finite and scale-independent result. Even though four quark-correlators are not strictly needed for the calculation of  susceptibilities, we will nevertheless calculate them, since they are needed to test the factorization hypothesis in the thermal case,
because we aim to determine their validity as order parameters,
and because they allow for a direct consistency check of our susceptibility calculation.

\subsection{ChPT thermal four-quark scalar correlators and condensates.}
\label{sec:fourq}

Here we will calculate the relevant four-quark scalar correlators, by taking the corresponding functional derivatives, as in Eq.~(\ref{fourquarkcorlight}), from the effective Lagrangian in Eq.~(\ref{zeff}) at a given order in ChPT. We also calculate the two-quark condensate  Eq.~\eqref{cond Z} to the same order, to check the  factorization hypothesis. We have:

\begin{equation}\label{qq Leff}
  \condtwo_T=  -\left\langle\frac{\delta \mathcal
    L_{eff}[s]}{\delta s_0(x)}+\frac{\delta \mathcal
    L_{eff}[s]}{\delta s_s(x)}\right\rangle_T\bigg\vert_{s=\mathcal M},
\end{equation}

\begin{eqnarray}\label{quarkcor Leff}
  \quarkcorT&=&\left\langle {\cal T}\left(\frac{\delta}{\delta s_0(x)}+\frac{\delta}{\delta s_s(x)}\right) \left(\frac{\delta}{\delta s_0(0)}+\frac{\delta}{\delta s_s(0)}\right)  \mathcal
  L_{eff}[s]\right\rangle_T\bigg\vert_{s={\cal M}}\delta(\tau)\delta^{(D-1)}(x)\nonumber\\
&+&\left\langle{\cal T}\left(\frac{\delta \mathcal
  L_{eff}[s]}{\delta s_0(x)}+\frac{\delta \mathcal
  L_{eff}[s]}{\delta s_s(x)}\right)\left(\frac{\delta \mathcal
  L_{eff}[s]}{\delta s_0(0)}+\frac{\delta \mathcal
  L_{eff}[s]}{\delta s_s(0)}\right)\right\rangle_T\bigg\vert_{s=\mathcal M}.
\end{eqnarray}

All our results can be expressed in terms of the leading order (free) thermal meson propagators $G_i^T(x)$, with $i=\pi$, $K$, $\eta$. Using standard finite-temperature methods, one can separate the $T=0$ part. The divergent contribution is contained in the $x=0$ and $T=0$ part. We follow the notation of \cite{Gerber:1988tt} for thermal functions:

\begin{eqnarray}
G_i^T(0)&=&G_i(0)+g_1(M_i,T),\nonumber \\
g_1(M,T)&=&\frac{1}{2\pi^2}\int_0^\infty dp \frac{p^2}{E_p} \frac{1}{e^{\beta
    E_p}-1},
\label{g1def}
\end{eqnarray}
with $E_p=\sqrt{p^2+M^2}$ and $G_i(0)=M_{0i}^{D-2}\Gamma\left[1-D/2 \right]/(4\pi)^{D/2}$, the $T=0$ divergent part in the dimensional regularization scheme, which we will use throughout this work. The  renormalization of  the LEC and the quark condensates up to NNLO in ChPT are the same as at $T=0$ and are discussed in detail in \cite{GomezNicola:2010tb}.

Now, from Eqs.~\eqref{qq Leff} and \eqref{quarkcor Leff}, using the
Lagrangians in \cite{Gasser:1984gg} and \cite{Bijnens:1999sh}, we obtain the
following results:
\begin{eqnarray}
\quarkcor_{T,\;NLO}\!\!&=& \!\!\condtwo_{T,NLO}^2,\qquad\label{Qsu3NLO}\\
\quarkcor_{T,\;NNLO}\!\!&=& \!\!\condtwo_{T,NNLO}^2+B_0^2 \left[24(12L_6+2L_8+H_2)\delta(\tau)\delta^{(D-1)}(x)+K^T(x)\right],
\label{Qsu3NNLO}
\end{eqnarray}
where the NLO and NNLO are $\Od(F^2)$ and $\Od(F^2)+\Od(F^0)$ in the ChPT counting respectively and $K^{T}(x)$ is the connected part of the four-meson correlator at leading order and finite temperature:
\begin{equation}\label{Ksu3}
K^T(x)=\langle {\cal T}\phi^a(x)\phi_a(x)\phi^b(0)\phi_b(0)\rangle_{T,LO}-\langle {\cal T}\phi^a(0)\phi_a(0)\rangle^2_{T,LO}=2\parent{3G^T_\pi(x)^2+4G^T_K(x)^2+G^T_\eta(x)^2}.
\end{equation}

We have expressed our results as a function of the square of the thermal quark
condensate $\condtwo_T$, whose explicit expressions
are given in Appendix \ref{app:renquark}, Eqs.~(\ref{condtwosu3nlo})-(\ref{condsnnlo}). As it happened in the $T=0$ case
\cite{GomezNicola:2010tb}, up to NLO the four-quark correlator is just equal
to the square of the quark condensate, but to NNLO the
connected one-loop contribution breaks such equality. For a detailed diagrammatic description of the different terms contributing to the four-quark correlator, we refer to \cite{GomezNicola:2010tb}, since the diagrams are the same at $T\neq 0$.

At this point, it is important to remark that we provide the full ChPT $SU(3)$ result for the two-quark condensates to NNLO including all the relevant meson interactions. Previous works at finite temperature only included the kaon and eta (and other massive states such as nucleons) as free fields in the partition function \cite{Gerber:1988tt}, which is reasonable for heavier particles when $T<<M_{K,\eta}$, since the heavy states are Boltzmann suppressed.  Therefore, apart for the study of factorization, our complete NNLO calculation of the condensate will allow to test the effect of including in the diagrams contact interactions containing strange particles.

Similarly to the previous analysis, we  calculate separately the light, strange and mixed four-quark correlators, which also factorize up to NLO in the product of the two-quark condensates, whereas up to NNLO we get:
\begin{eqnarray}
\quarkcorlT \!\!&=& \!\!\langle\parent{\bar q q}_l\rangle_{T}^2+B_0^2
\left[16(8L_6+2L_8+H_2)\delta(\tau)\delta^{(D-1)}(x)\right.\nonumber\\
  & &\quad\quad\quad\quad\quad\;\;\;\left.+6G^T_\pi(x)^2+2G^T_K(x)^2+\frac{2}{9}G_\eta^{T}(x)^2\right] +\Od\left(\frac{1}{F^2}\right),\qquad\label{Qlsu3}\\
\langle {\cal T}(\bar s s)(x) (\bar s s)(0) \rangle_T \!\! &=& \!\!\condsT^2+B_0^2 \left[8(4L_6+2L_8+H_2)\delta(\tau)\delta^{(D-1)}(x)+2G_K^T(x)^2+\frac{8}{9}G_\eta^T(x)^2\right] +\Od\left(\frac{1}{F^2}\right),\qquad\label{Qssu3}\\
\langle {\cal T}(\bar q q)_l(x) (\bar s s)(0) \rangle_T \!\! &=& \!\!\langle \parent{\bar q q}_l \rangle_T\condsT+B_0^2\left[64L_6 \delta(\tau)\delta^{(D-1)}(x)+2G_K^T(x)^2+\frac{4}{9}G_\eta^T(x)^2\right]+\Od\left(\frac{1}{F^2}\right).
\label{Qqsu3}
\end{eqnarray}

 \subsection{Factorization breaking at finite temperature}\label{sec:factorization}

As discussed in the introduction, scalar condensates play a relevant role in QCD, since they are
directly related to  vacuum properties. Attending only to their symmetry transformation properties,  quark condensates of arbitrary order
$\condn$ should behave similarly
as the two-quark condensate under chiral restoration, since they transform as isoscalars and are built out of chiral non-invariant operators with
the vacuum quantum numbers. Actually, it is well known that such quark condensates appear directly in QCD sum
rules, through the  Operator Product Expansion
(OPE) approach \cite{Shifman:1978bx}.
In that framework, the following hypothesis of factorization or
vacuum saturation is customarily made:
\begin{equation}
\condfour=\left(1-\frac{1}{4N_c N_f}\right)\condtwo^2,
\label{facthyp}
\end{equation}
and similarly for other condensates, which  in the large-$N_c$
limit  simply reduces to $\condfour=\condtwo^2$. The
second term between brackets in Eq.~(\ref{facthyp}) comes from the exchange of indices (including color)
between the first and second $\bar q q$ operators.

The use of the factorization hypothesis has been a much debated tool in order
to estimate the size of higher order condensates in the OPE.
Let us remark that in \cite{GomezNicola:2010tb} we have shown that factorization of the four-quark condensate does not hold within the model-independent QCD low-energy regime provided by ChPT, at $T=0$. Actually, once the two-quark condensate is renormalized at
a given order, factorization breaking terms are divergent and dependent on the low-energy scale. Therefore, the scalar four-quark condensate is not even a low-energy observable at $T=0$.
 This factorization breaking  is consistent with previous RG analysis \cite{Narison:1983kn}.
This hypothesis, also known as ``Vacuum saturation'' has also been studied within the sum rule approach and found
not
to work (see, for instance \cite{Dominguez:2006ct}). Nevertheless, in \cite{GomezNicola:2010tb} we found that factorization holds formally if the large $N_c$ limit
is taken before renormalization,
since factorization breaking terms are $\Od(N_c^{-2})$ suppressed.

In this paper we extend the analysis performed in \cite{GomezNicola:2010tb} to the $T\neq 0$ case, since, among other reasons, this can shed light on the use of the four-quark condensate $\condfourT$ as an order parameter of the chiral transition. In particular, in case factorization holds, the four-quark condensate should behave as an order parameter, melting at the same critical temperature as the two-quark condensate.
We obtain readily the factorization properties of
scalar four-quark condensates by setting   $x\rightarrow 0$
in our results in section \ref{sec:fourq}. It is clear then that,
at finite temperature, factorization does not hold either.
 In fact, since $\delta^{(D-1)}(0)$ vanishes identically in dimensional
 regularization \cite{Leibbrandt:1975dj}, we get from Eqs.\eqref{cond4qdef} and \eqref{Qsu3NNLO}:
\begin{equation}\label{factbreakingsu3}
  \frac{{\condfourT}}{{\condtwoT}^2}=1+\frac{2}{9F^4}\parent{3G^T_\pi(0)^2+4G^T_K(0)^2+G^T_\eta(0)^2}+{\cal O}(1/F^6).
\end{equation}

We can calculate again the light, strange and mixed four-quark cases separately in $SU(3)$, namely:
\begin{equation}
\frac{\langle\left(\bar q q\right)_l^2\rangle_T}{{\condtwolT}^2}=1+\frac{1}{2F^4}\parent{3G^T_\pi(0)^2+G^T_K(0)^2+\frac{1}{9}G^T_\eta(0)^2}+{\cal O}(1/F^6),
\label{nonfac4ql}
\end{equation}

\begin{equation}
\frac{\langle\left(\bar s s\right)^2\rangle_T}{\langle
  \bar s s\rangle^2_T}=1+\frac{2}{F^4}\parent{G^T_K(0)^2+\frac{4}{9}G^T_\eta(0)^2}+{\cal O}(1/F^6),
\end{equation}

\begin{equation}
\frac{\langle\left(\bar q q\right)_l(\bar ss)\rangle_T}{\langle (\bar q q)_l\rangle_{T}\langle ss\rangle_T}=1+\frac{1}{F^4}\parent{G^T_K(0)^2+\frac{2}{9}G^T_\eta(0)^2}+{\cal O}(1/F^6),
\label{nonfac4qmixed}
\end{equation}

In view of these results, several remarks are in order: First, the factorization breaking terms at $T\neq0$ are divergent and independent of the LEC, as for $T=0$ \cite{GomezNicola:2010tb}, which was expected since the finite temperature result for the four-quark condensates to this order is just obtained by replacing $G_i(0)\rightarrow G_i^T(0)$. Since the quark NNLO condensates are rendered finite by the renormalization of the $\Od(p^4)$ and $\Od(p^6)$ LEC  \cite{GomezNicola:2010tb}, this means that the four-condensate to this order is divergent, also at $T\neq 0$. Even subtracting the $T=0$ factorization breaking term, the result is still divergent, since, from Eq.~\eqref{g1def}, there are terms in Eq.~\eqref{factbreakingsu3} proportional to $G_i(0) g_1(M_i,T)$. Recall that, as mentioned at the beginning of this section, the four-quark condensates cannot be expressed only in terms of
mass derivatives of the free energy density, unlike the quark condensates.

Second, in the chiral limit, and contrary to the zero temperature case,
for $T\neq0$ the
factorization breaking terms $G^T_i(0)$ in Eq.~\eqref{factbreakingsu3} are finite and do not
vanish. Let us recall that  in
dimensional regularization the $T=0$ propagators $G_i(0)=0$ when $M_i=0$, since they
are proportional to $M_i^2$. However, the thermal part $g_1(M_i,T)$ is finite and non vanishing
in the chiral limit. Thus, the chiral limit is the only case for which the four-quark condensate can be considered an order parameter. Actually, we have checked that $\langle\left(\bar q q\right)_l^2\rangle_T/\langle\left(\bar q q\right)_l^2\rangle_0$ follows the same temperature melting behavior as $\condtwol_T/\condtwol_0$ for the  temperatures where ChPT is reliable. Note that chiral symmetry restoration takes place formally only in the chiral limit.

Third, for $T\neq0$ factorization holds formally
in the $N_c\rightarrow\infty$ limit,
as it happened for the $T=0$ case \cite{GomezNicola:2010tb},
since $F^2=\Od(N_c)$ and therefore factorization breaking terms are once again $\Od(N_c^{-2})$ suppressed.

Partial results on the non-factorization of the thermal four quark condensates in certain approximations also exist in the literature. For instance, using the soft-pion and chiral
limits it has been shown in \cite{Eletsky:1992xd} that if one assumes
factorization at zero temperature, it will be
spoiled  by the lowest order thermal corrections.
Our results support those in \cite{Eletsky:1992xd}
since, on the one hand, we have proved factorization  for  $T=0$  in the chiral limit, and, on the other hand, we have also found that factorization breaks due to $T\neq0$ contributions.

In addition, an analysis of medium effects indicates that factorization is expected to be
broken by $1/N_c^2$ suppressed contributions due to particles with the same quantum
numbers as the operator under consideration \cite{Leupold:2005eq}.
Hence, we formally agree with  \cite{Leupold:2005eq}, since the leading term in $1/N_c$ for the scalar four-quark condensate in the pure pion gas should factorize due to the absence of crossed terms $\langle\pi\vert \bar q q \vert 0\rangle$.
Nevertheless, by dropping these $1/N_c^2$ subleading breaking terms,
some combinations of four-quark condensates, which include
that in Eq.\eqref{nonfac4ql} above, have been proposed in \cite{Leupold:2006ih} as order parameters.
However, as we have just commented, even if these divergences are formally $1/N_c$ suppressed,
the four-quark condensates in Eq.\eqref{nonfac4ql}
are actually divergent.
Therefore, for finite $N_c$,
the use of the order parameters proposed in \cite{Leupold:2006ih}
would require checking the cancellation of
the divergences in the combinations
suggested in \cite{Leupold:2006ih}.

Thus, one of the main conclusions of the present work is that
the four-quark condensates in Eqs.\eqref{factbreakingsu3} to \eqref{nonfac4qmixed}, cannot be used as order parameters of the chiral phase transition, because they are divergent. Their use  in previous works
is
valid only formally for $N_c\rightarrow\infty$ or the chiral limit, but not in the physical case. We point out that a breaking of four-quark condensate factorization at finite temperature has been reported also in \cite{Johnson:1999ps} within the framework of QCD sum rules.

\subsection{Scalar susceptibilities within ChPT}
\label{sec:sus}

We now turn to the ChPT evaluation of the different susceptibilities defined in section \ref{sec:gendef}. As detailed in Eq.~(\ref{chil4q}) and  Eqs.~(\ref{chis4q}) to (\ref{chitot4q}), the scalar susceptibilities can be calculated either as a mass derivative of a quark condensate or from the corresponding four-quark correlators. Since we have available both the condensates (displayed in Appendix \ref{app:renquark}) and the four-quark correlators given in section \ref{sec:fourq}, we can obtain the ChPT susceptibilities in both ways, thus checking our results for the four-quark correlators. Recall that precisely the additional factorization-breaking like terms in the NNLO correlators in Eqs.~(\ref{Qsu3NNLO}) and (\ref{Qlsu3})-(\ref{Qqsu3})  give nonzero susceptibilities of chiral order $\Od(1)$. That order correspond to the derivatives of the NLO condensates in Appendix \ref{app:renquark}, since the $\Od(F^2)$ is mass independent. If we calculate the susceptibilities from the four-quark correlators, we check again that the $\Od(F^2)$ vanishes (NLO correlators) while from the NNLO correlators,  taking into account that
\begin{equation}
\int_T d^4 x \left[G_i^T (x)\right]^2=-\frac{d}{d M^2_i} G_i^T (0),
\end{equation}
we readily obtain the different ChPT scalar susceptibilities to $\Od(1)$ in the chiral power counting:

\begin{eqnarray}
\chi (T)&=&B_0^2\left[ 24\parent{12L^r_6(\mu)+2L^r_8(\mu)+H^r_2(\mu)}-12\nu_\pi-16\nu_K-4\nu_\eta\right.\nonumber\\
&+&\left.6g_2(M_\pi,T)+8g_2(M_K,T)
+2g_2(M_\eta,T)\right]+\Od\left(\frac{1}{F^2}\right),
\label{suscsu3}\\
\chi_{l}(T)&=&B_0^2\left[16\left(8L_6^r(\mu)+2L_8^r(\mu)+H_2^r(\mu)\right)-12\nu_\pi-4\nu_K-\frac{4}{9}\nu_\eta\right.\nonumber\\
&+&\left.6g_2(M_\pi,T)+2g_2(M_K,T)
+\frac{2}{9}g_2(M_\eta,T)\right]+\Od\left(\frac{1}{F^2}\right),
\label{susclightsu3}\\
\chi_{s}(T)&=&B_0^2\left[8\left(4L_6^r(\mu)+2 L_8^r(\mu)+H_2^r(\mu)\right)-4\nu_K-\frac{16}{9}\nu_\eta+2g_2(M_K,T)+\frac{8}{9}g_2(M_\eta,T)\right]+\Od\left(\frac{1}{F^2}\right)
\label{suscs}\\
\chi_{ls} (T)&=&B_0^2\left[64L_6^r(\mu)-4\nu_K-\frac{8}{9}\nu_\eta+2g_2(M_K,T)+\frac{4}{9}g_2(M_\eta,T)\right]+\Od\left(\frac{1}{F^2}\right),
\label{suscls}
\end{eqnarray}
with:
\begin{eqnarray}
\nu_i&=&\frac{1}{32\pi^2}\left(1+\log\frac{M_{0i}^2}{\mu^2}\right),\label{nudef}\\
g_2(M,T)&=&-\frac{dg_1(M,T)}{dM^2}=\frac{1}{4\pi^2}\int_0^\infty dp \frac{1}{E_p} \frac{1}{e^{\beta E_p}-1}.
\label{g2def}
\end{eqnarray}

A further check is that the results for the different susceptibilities are finite and independent of the low-energy scale $\mu$, unlike the four-quark condensates, with the same renormalization of the low-energy constants, provided in \cite{GomezNicola:2010tb}, which ensures that quark condensates are also finite and scale independent to that order.

 Note that,
in the $T\ll M_{K}$ regime,
the thermal pion loops, i.e. $g_2(M_\pi,T)$,
dominate over other particle thermal loop contributions,
which are Boltzmann suppressed.
An even more interesting regime is $M_K\gg T\gg M_\pi$, because
it is related to the critical behavior \cite{Smilga:1995qf}.
Within our approach we cannot reach high temperatures but we can study the near chiral limit,
where these functions behave as $g_2(M_\pi,T)\simeq T/(8\pi M_\pi)$.
We will remark this expected linear behavior when we plot our results below.

Note that pion terms show up in the
light susceptibility $\chi_l$ but not in the strange and mixed ones $\chi_s,\chi_{ls}$ which
are therefore subdominant compared to the light one at temperatures below the transition.
We have explicitly checked that in the limit $m_s\rightarrow\infty$ $(M_{K,\eta}\rightarrow\infty)$,  the $SU(3)$
light susceptibility  reduces to the $SU(2)$ result in Eq.~\eqref{su2sus} in Appendix \ref{sec:su2res},
once the identification between $SU(2)$ $l_i^r$ and $L_i^r$ LEC is made, as given in \cite{Gasser:1984gg}.

The thermal scaling of the light susceptibility near the critical  point $(m_{u,d},T)\rightarrow (0^+,T_c)$ reveals important features about the nature of the phase transition and is the subject of detailed analysis in lattice simulations \cite{Ejiri:2009ac}. The leading pion mass dependence from Eq.(\ref{susclightsu3}) near the chiral limit  is $\chi^{IR}_{l,T=0}/B_0^2\sim -\frac{3}{8\pi^2}\log (M_\pi^2/\mu^2)$  and $(\chi_{l,T}-\chi_{l,T=0})^{IR}/B_0^2\sim \frac{3T}{4\pi M_\pi}$ for $M_\pi\ll T \ll M_{K,\eta}$. This leading behavior was already found in \cite{Smilga:1995qf}. In addition, our results for the light susceptibility are consistent with
a recent
and model-independent ChPT analysis \cite{Nicola:2011gq} of the mass,
temperature and flavor dependence of the light
susceptibilities, where a separate analysis of the quark connected and disconnected contributions was provided.

After providing the analytic expressions for the susceptibilities, and addressing some formal aspects, let us discuss their phenomenology in connection with that of the quark condensate.

\section{Phenomenological results in ChPT}
\label{sec:pheno}

\subsection{Higher order meson interactions and the quark condensate}
\label{sec:effecthigh}

 As we have stated in the previous section, we provide in this work for the first time the full NNLO ChPT SU(3)  $\condtwo_T$ results, which are model-independent and given in Appendix~\ref{app:renquark} (the SU(2) one is also given in that Appendix). This result includes all the  meson-meson interactions for $\pi,K,\eta$ up to that order. At NLO, only tadpole contributions proportional to $g_1(M_i,T)$ appear, which is equivalent to considering an ideal gas made of the mesonic components. Previous studies of the condensate \cite{Gerber:1988tt} actually considered the contribution of the heavier $K,\eta$ states as free, while keeping higher orders in pure pion interactions. Here we will extrapolate our results to the critical point and make an estimate of the effect of considering interactions involving those strange degrees of freedom, by comparing with the NLO and with the pure $SU(2)$ result in which kaons and eta have been decoupled.

In order to study the thermal effects it is customary to normalize
the quark condensate  to its $T=0$ value and so we will do in what follows.
An additional advantage of this normalization is that the ratio is QCD Renormalization Group (RG) independent, because the $B_0$ global factors cancel. Let us note that, to NLO no LEC appear in $\condtwo_T/\condtwo_0$.
 To NNLO, only the $\Od(p^4)$ LEC do appear.
For the ChPT plots that we will present next,
we will use the $SU(3)$ set of $L_i^r$ LEC from \cite{Amoros:2001cp} and their error bands.
  An important remark is that, by definition, the $H_2^r$ constant cannot
be fixed directly from experimental meson data. This constant appears at
NNLO and we  estimate it  as $H_2^r=2L_8^r$
 as suggested in \cite{Ecker:1988te} using scalar resonance saturation, which was also used in
\cite{Amoros:2001cp}, where it was remarked that $H_2^r$ has a rather small impact in the value of the non-strange condensates.
Let us remark that there are newer determinations of all the $L_i^r$
LEC in \cite{Bijnens:2011tb} and of
$H_2^r$ and $L_8^r$ in \cite{Bordes:2012ud}, although the best values for $L_8^r$ from these two recent references are more than two standard deviations away. We have thus preferred to stick to the older reference \cite{Amoros:2001cp}, which has a value of $L_8^r$ lying
between the two new ones, and also provides the complete set of LECs.
In addition, as it is customary, at each given order
we use $F=F_\pi(1+{\cal O}(M_i^2/F^2))$, $M_{0i}^2=M_i^2(1+{\cal O}(M_i^2/F^2))$ \cite{Gasser:1984gg} taking the physical values for $F_\pi$ and $M_i^2$ and including the corrections in the next order.
\begin{figure}
\includegraphics[scale=0.5]{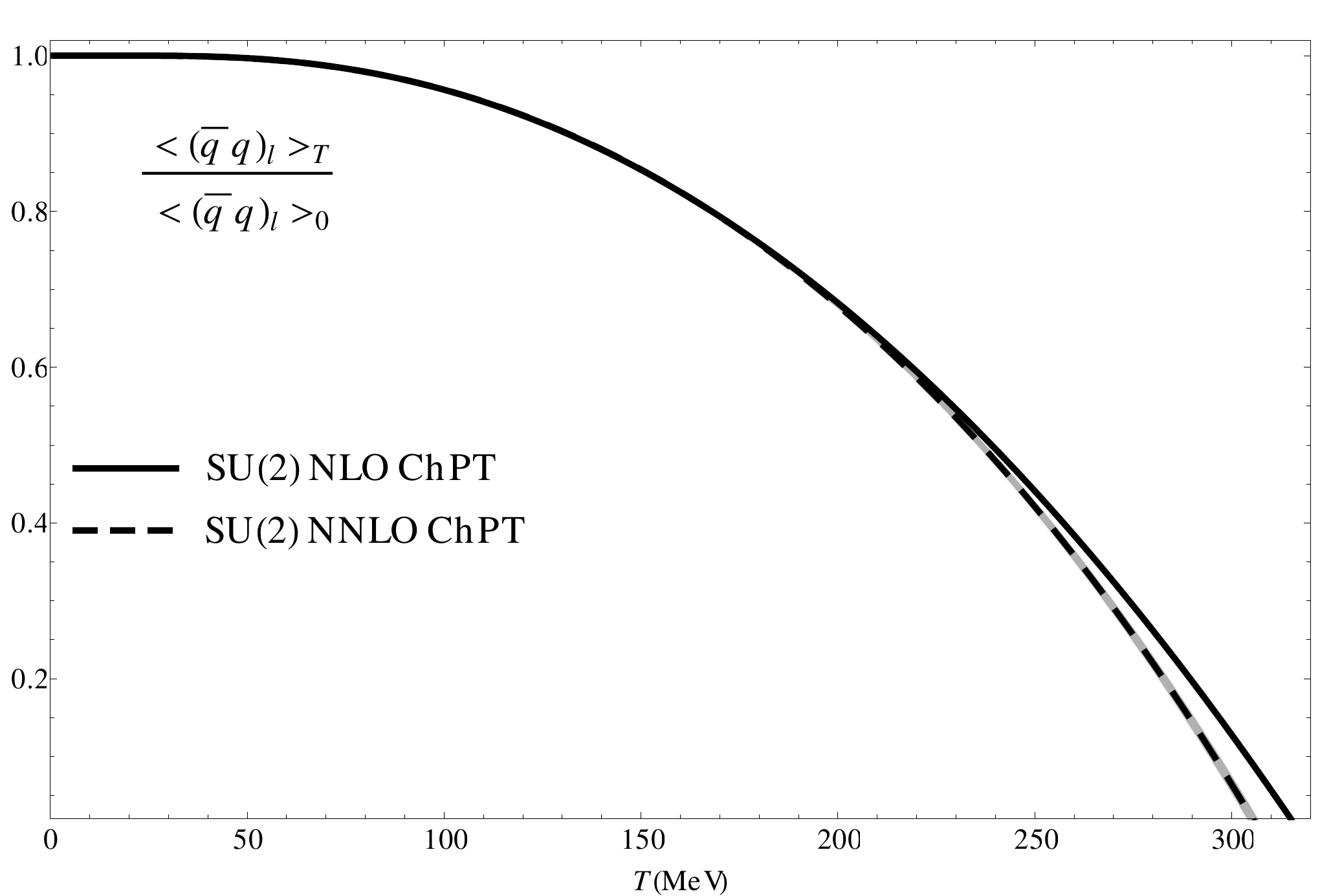}
\includegraphics[scale=0.5]{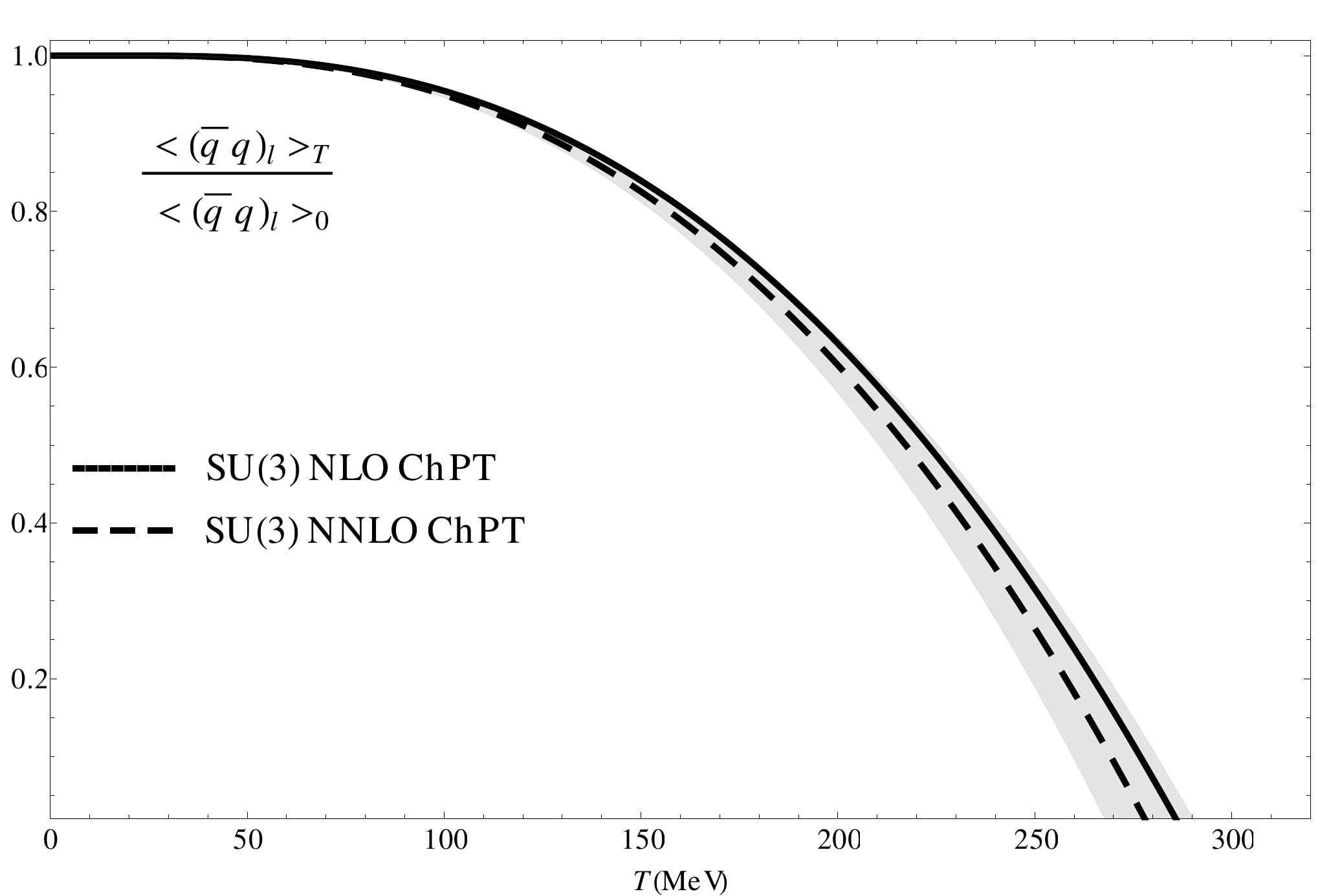}
 \caption{\rm \label{fig:condchpt} Non-strange quark condensate in ChPT as a function of temperature in the different cases explained in the main text.}
\end{figure}

Now,  to comment on the relative size of different effects under discussion, we are going to extrapolate our results up to temperatures beyond the strict applicability limit of ChPT, typically estimated around $T=150\,$MeV \cite{Gerber:1988tt}. The reason is that it is much easier to explain
the changes due to different effects in the curves of the condensate ratios, by comparing the points where their corresponding extrapolation vanish. We will refer to this point as the ``critical temperature'', $T_c$, in the clear understanding that this is just for the sake of comparison between curves, since the particular value of this ``extrapolated'' temperature is just a very crude extrapolation and only
the low temperature part is reliable and model independent. We will proceed similarly in section \ref{sec:vir} about the virial expansion.

In order to study the restoration of the spontaneously broken chiral symmetry it is customary to use the non-strange condensate. The reason not to include the thermal strange quark condensate is that it has a very slow decrease as $T$ increases, because its evolution is dominated by the larger $m_s$ mass, which is an explicit and not a spontaneous breaking.
Thus, in the upper panel of Fig.~\ref{fig:condchpt}
we show the thermal dependence of the
normalized non-strange condensate within NLO and NNLO $SU(2)$ ChPT,
which can be compared with the $SU(3)$ version in the lower panel.
The gray bands surrounding the NNLO calculation cover the uncertainties in the LEC. We see that the NNLO correction is relatively small compared to the NLO one below $T_c$.  The approximate temperature
 at which the $SU(2)$ condensate vanishes is
consistent with \cite{Gerber:1988tt}, when we compare to the same order of approximation, taking into account that we use a more recent
set of LECs. The reduction in $T_c$ from the $SU(2)$ to the $SU(3)$ case can be interpreted as a
``paramagnetic" effect coming from the increase in entropy (and therefore disorder) due to the addition of the strange degrees of freedom to the system.
To NLO, the reduction on $T_c$ from $SU(2)$ to $SU(3)$ is of 30 MeV
whereas at NNLO is of $28\pm12\,$MeV. This is the behavior expected when adding more degrees of freedom as they become more relevant near the phase transition, and it actually provides a natural explanation to the smaller values of $T_c$ obtained in approaches such as the HRG which take into account all the relevant (free) degrees of freedom. In the $SU(2)$ analysis we also observe a change in the value of $T_c$ due to considering interacting pions (i.e., from NLO to NNLO) of about $\Delta T_c\sim -10\,$MeV. However, the result is inconclusive for
the $SU(3)$ case due to the uncertainties from the LEC dependence at NNLO.

\subsection{ChPT susceptibility}

\begin{figure}
\includegraphics[scale=.43]{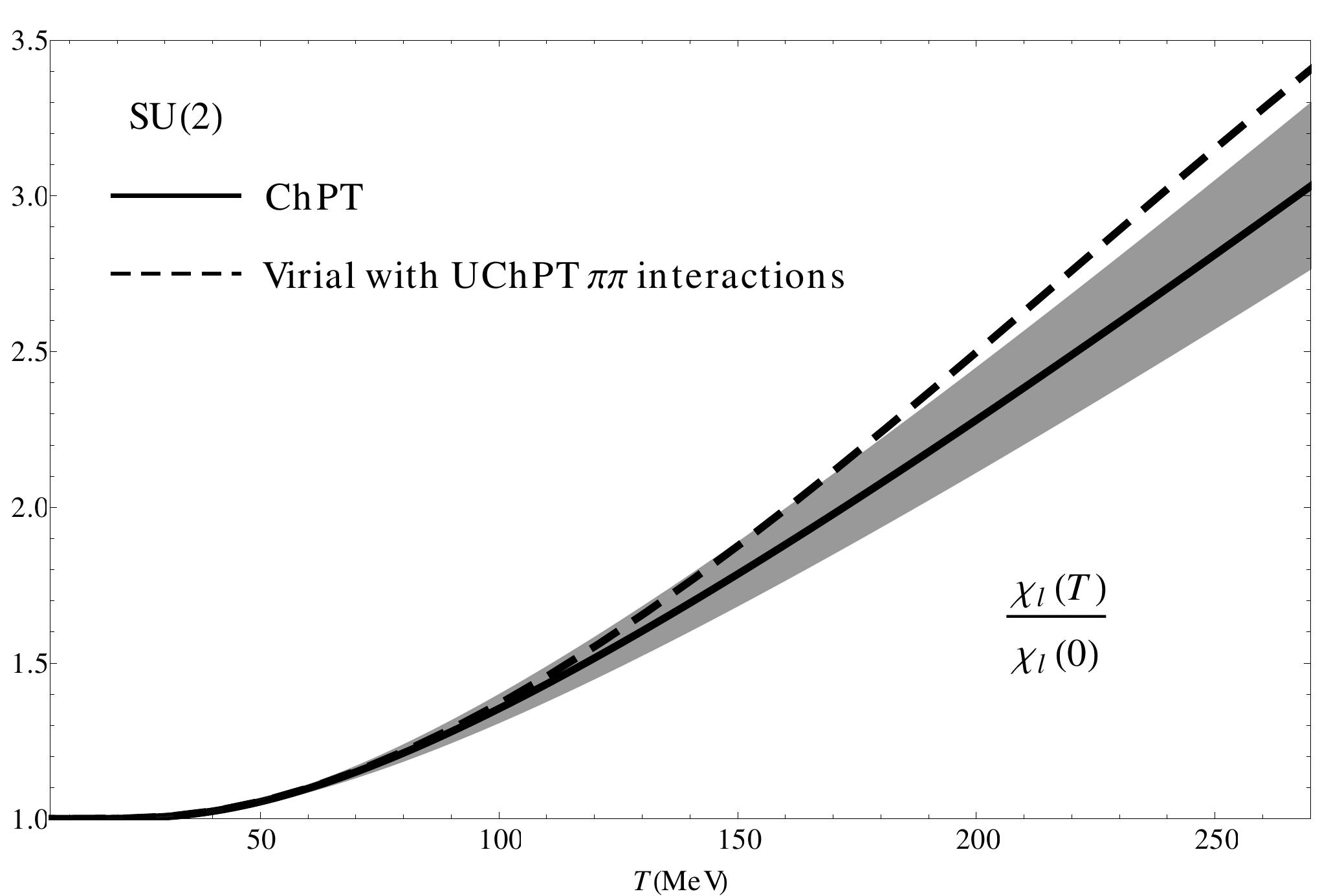}
\includegraphics[scale=.43]{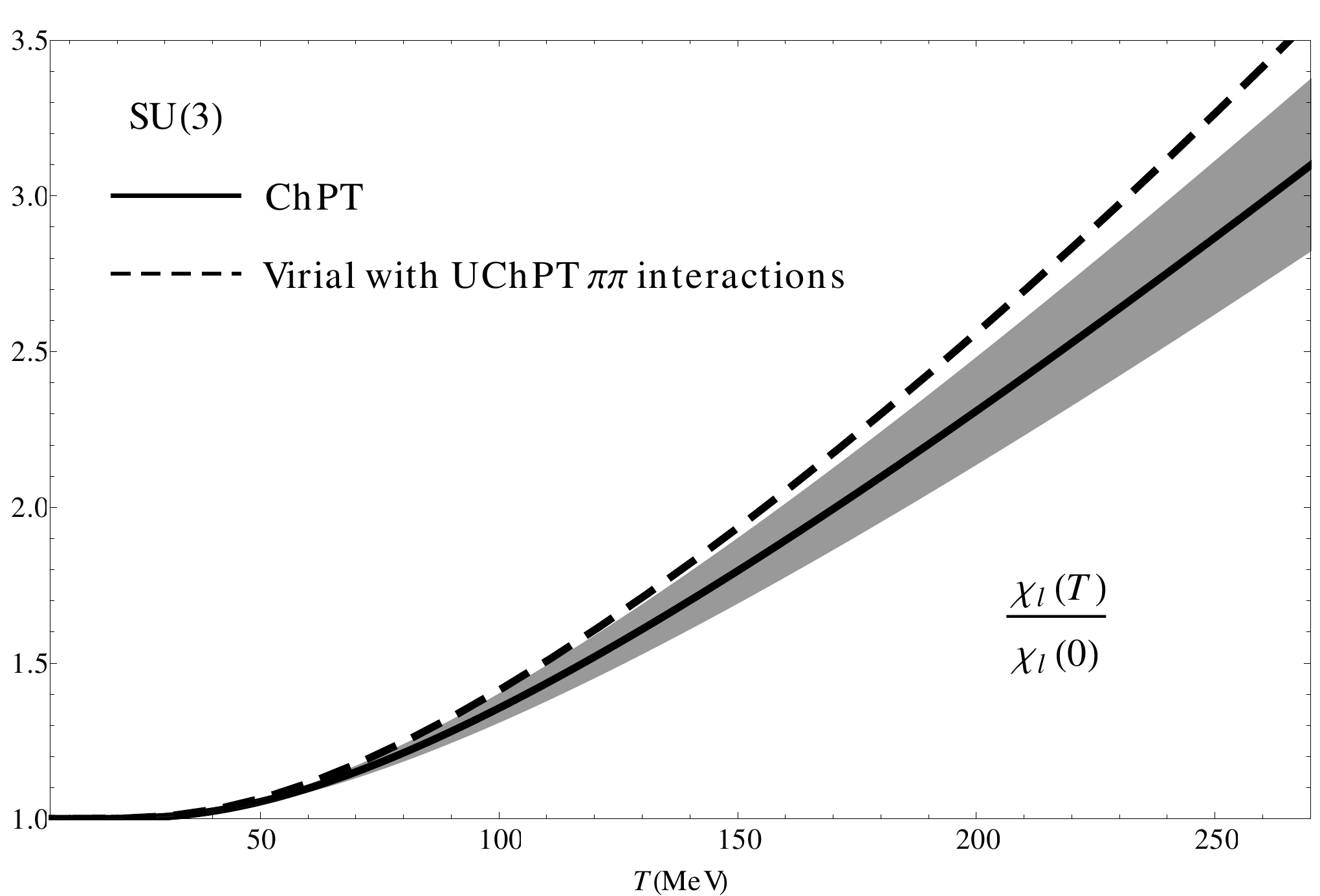}
 \caption{\rm \label{fig:suscChPT} Non-strange scalar susceptibility as a function of temperature for the $SU(2)$ and $SU(3)$ cases in ChPT at NNLO. The uncertainty bands, due to the uncertainties in the LEC, have similar size for cases cases. The curves for the virial approach give a crude estimate of several sources of systematic uncertainties discussed in the text.}
\end{figure}

We can also analyze the thermal evolution towards chiral restoration by studying the chiral susceptibility thermal dependence. In Fig.~\ref{fig:suscChPT} we plot the ChPT $\chi_{l,T}$ result obtained for the $SU(2)$ and $SU(3)$ cases
using Eqs.~\eqref{susclightsu3} and \eqref{su2sus}, respectively, normalized to their $T=0$ values and for the same set of parameters than the quark condensate in Fig.~\ref{fig:condchpt}.
Note that we plot the ratio of the thermal to $T=0$ susceptibilities, which, once again, is QCD RG independent due to the cancellation of the $B_0^2$ factors.
We do not expect our low-energy analysis to reproduce the dramatic growth just below the critical point for $\chi_l$. As it also happens for the quark condensate, we expect only to reproduce reasonably the low $T$ side of the critical curves. We will provide a quantitative comparison with lattice values in section \ref{sec:latcom} below.  As we have discussed above, the temperature evolution is governed  by the $T$-increasing functions $g_2$, so that, as we see in Fig.~\ref{fig:suscChPT}, the ChPT growth is roughly linear for temperatures high enough, according to the behavior in the chiral limit.  In fact, although we have taken into account the full contribution of $\pi,K,\eta$ loops in the $SU(3)$ massive case, the kaon and eta thermal contributions provide a weak dependence, so that the $SU(2)$ and $SU(3)$ curves are very similar to one another up to the critical point, unlike  the  ``paramagnetic" shift for the case of  the quark condensate. In fact, although in both cases the $SU(2)$ and $SU(3)$ differ by terms of order $e^{-M_{K,\eta}/T}$, the thermal dependence with $\tau=T/M_{K,\eta}$ is much softer in the susceptibility, since it comes from $g_2(M_{K,\eta},T)\sim \sqrt{\tau}e^{-1/\tau}$, than in the condensate where $g_1(M_{K,\eta},T)\sim T^2 (1/\sqrt{\tau})e^{-1/\tau}$ for $\tau\ll 1$.   Again, the strange quark susceptibility is not showed, $\chi_s/\chi_0$ being also increasing but remaining very close to unity for the range of temperatures showed in Fig.~\ref{fig:suscChPT}.

\section{The virial approach}
\label{sec:vir}

In previous sections we have been able to estimate the effect of the uncertainties in the LEC. However, ChPT is a perturbative expansion that, when truncated, neglects higher order corrections and cannot reproduce resonances. This is a first motivation to use the virial expansion, because it allows for a simple implementation of unitarized ChPT, which includes the numerically relevant higher order effects. An additional motivation to use the virial approach is that the  interaction part can be clearly identified and described realistically. Both features will allow us to establish a consistent comparison with our previous standard ChPT approach.

The virial expansion is a simple and successful approach already applied to
describe many thermodynamical properties of dilute gases made of interacting
pions \cite{Dashen,Welke:1990za,Dobado:1998tv}  and other hadrons
\cite{Gerber:1988tt,Venugopalan:1992hy,Nyffeler:1993iz,Pelaez:2002xf,GarciaMartin:2006jj}. A similar approach for the calculation of the quark condensate
was suggested in \cite{Chanfray:1996bd}. For most thermal observables, it is
enough to know the $T=0$ scattering phase shifts of the particles that compose
the gas. In principle these phase shifts could be taken from experiment avoiding any model dependence,
so that it would not be necessary to go through the technicalities of finite
temperature field theory. However, if  one is interested in
chiral symmetry restoration, and hence in scalar
susceptibilities and quark condensates,  one needs a model-independent
theoretical description of the phase shifts in order to obtain their quark mass dependence, which cannot be obtained directly from
experiment. Using directly one-loop ChPT for scattering lengths provides a remarkable description of the low energy hadronic
interactions and should be accurate enough at very low temperatures \cite{Gerber:1988tt}. However, for temperatures further away from the threshold region, a more precise description of the scattering is needed, which in particular takes into account the loss of unitarity, and the absence of resonances, in the
pure ChPT expansion. For that purpose, we will make use of the so called unitarized ChPT, at the expense of loosing the systematic ordering of the effective approach.

The thermodynamics of a system of hadrons
is encoded in the free energy
density $z=\epsilon_0-P$, where $\epsilon_0=z^{T=0}$ and $P$ is the pressure.
In the present work,
we are interested in a multi-component interacting relativistic gas made of pions, kaons
and etas in thermal and chemical equilibrium, so the pressure only depends on temperature $T$. Thus, the second order relativistic virial expansion of the pressure reads \cite{Dashen,Gerber:1988tt,Welke:1990za,Dobado:1998tv}:
\begin{equation}\label{virial pressure}
   \beta P=\sum_{i}{\parent{B_i^{(1)}\xi_i+B_{i}^{(2)}\xi_i^2+\sum_{j\geq i}{B_{ij}^{\mathrm{int}}\xi_i\xi_j}}},
\end{equation}
where $i=\pi,K,\eta$,  the fugacities $\xi_i=e^{-\beta M_i}$, $M_i$ is the mass of the $i$ species and the $B_i$ and $B_{ij}$ are the virial coefficients for the gas.
Expanding up to the second order in $\xi_i$ means that we only consider binary interactions. The coefficients
 \begin{equation} \label{virial free coefficients}
B_i^{(n)}=\frac{g_i}{2\pi^2 n}\int_0^\infty{dp\,p^2 e^{-n\beta(\sqrt{p^2+M_i^2}-M_i)}},
\end{equation}
where the degeneracy is $g_i=3,4,1$ for $\pi, K, \eta$ respectively, correspond simply to the virial expansion for a free gas
\begin{equation}\label{virial free pressure}
   \beta P_{\mathrm{free}}=-\sum_{i}{\frac{g_i}{2\pi^2}\int_0^\infty{dp\,p^2 \log\left[1-e^{-\beta(\sqrt{p^2+M_i^2})}\right]}}.
\end{equation}

The above free result is nothing but the HRG approach mentioned in the Introduction,
when considering only the lightest hadrons $\pi$, $K$, $\eta$. Thus, the virial expansion provides naturally the corrections to the HRG due to interactions, which appear through the $S$-matrix. For the meson-meson interactions, relevant for this work, this can be recast in terms of the elastic scattering phase shifts \cite{Dashen,Gerber:1988tt,Welke:1990za,Dobado:1998tv}. In this way, we can write
 \begin{equation}\label{virial interaction coefficients}
B^{\mathrm{int}}_{ij}=\frac{\xi_i^{-1}\xi_j^{-1}}{2\pi^3}\int_{M_i+M_j}^\infty{dE\,E^2K_1\left(E/T\right)\Delta_{ij}(E)}
\end{equation}
where $K_1$ is the first modified Bessel function and
 \begin{equation}\label{Deltaphases}
\Delta_{ij}(E)=\sum_{I,J,S}(2I+1)(2J+1)\,\delta_{I,J,S}^{ij}(E),
\end{equation}
where the $\delta^{ij}_{I,J,S}$ are the $ij\rightarrow ij$ elastic scattering phase shifts (chosen so that $\delta=0$ at threshold $E_{th}=M_i+M_j$) of a state $ij$  with quantum number $I,J,S$ (isospin, angular momentum and strangeness), that we will explain below.
The virial expansion  breaks down typically where the dilute gas expansion does, for $T\sim 200-250$ MeV \cite{Dobado:1998tv}.
In that regime $\xi_\pi \gg \xi_K \sim \xi_\eta$ so that the density of higher mass states an their interactions are Boltzmann suppressed with respect to pions. Hence, for our purposes of estimating systematic uncertainties, the $ij=\pi K$ and $\pi \eta$
states can be neglected against the $\pi\pi$ interactions and  we can drop the S index.

From Eq.~\eqref{cond z} we can then express the quark condensate as:
\begin{equation}\label{qqvirial0}
  \langle(\bar{q}q)_\alpha\rangle_T = \frac{\partial z}{\partial m_{\alpha}}=\langle 0\vert(\bar{q}q)_\alpha\vert 0\rangle-\frac{\partial P}{\partial m_\alpha},
\end{equation}
with $\alpha=l,s$, $m_l=m$ and  $\langle 0\vert(\bar{q}q)_\alpha\vert 0\rangle=\langle(\bar{q}q)_\alpha\rangle_{T=0}=\partial \epsilon_0/\partial m_\alpha$. We emphasize again that in order to calculate Eq.~\eqref{qqvirial0} we need the dependence of $\delta(E)$ on the quark masses as well as the vacuum expectation value. For that information we turn to ChPT in order to translate Eq.~\eqref{qqvirial0} in terms of physical meson masses:

\begin{equation}\label{qqvirial}
  \langle(\bar{q}q)_\alpha\rangle_T = \langle 0\vert(\bar{q}q)_\alpha\vert 0\rangle\parent{1+\sum_{i}{\frac{c_i^{\alpha}}{2M_iF^2}\frac{\partial P}{\partial M_i}}},
\end{equation}
with:
\begin{equation}\label{c's}
c_i^{\alpha}=-F^2\frac{\partial M_i^2}{\partial m_{q_\alpha}}\ematriz{0}{(\bar{q}q)_\alpha}{0}^{-1},
\end{equation}
for which we will take the one-loop ChPT expressions  from \cite{Gasser:1984gg,Pelaez:2002xf}. Since they depend on the ChPT LEC,
we take the same values as in previous sections, obtaining:
\begin{equation}
c_\pi^{\bar q q}=1.02,\qquad  c_K^{\bar q q}=0.59,\qquad c_\eta^{\bar q q}=0.52.\label{eq:cs}
\end{equation}
These numerical values are almost identical to those obtained in \cite{Pelaez:2002xf}.

The behavior of the light and strange condensates within the virial expansion has been analyzed in detail for the meson gas in \cite{Welke:1990za,Venugopalan:1992hy,Dobado:1998tv,Pelaez:2002xf} and also including baryon interactions in \cite{GarciaMartin:2006jj}.
In this paper we are interested mostly in four-quark condensates and susceptibilities.
The four-quark condensates, e.g  Eq.~(\ref{fourquarkcorlight}), cannot be obtained directly from the pressure as mass derivatives and hence the virial approach cannot give further information about them. The susceptibilities, defined in section \ref{sec:gendef} can be calculated taking one more mass derivative. Hence,

\begin{equation}
 \chi_{\alpha\beta}(T) =-\frac{\partial^2 z}{\partial
   m_\alpha\partial m_\beta}=\chi_{\alpha\beta}(0)+\frac{\partial^2 P}{\partial
   m_\alpha\partial m_\beta},
\end{equation}
and again translating it in terms of meson masses:
\begin{equation}\label{chivirial}
  \chi_{\alpha\beta}(T)=\chi_{\alpha\beta} (0)\left[1+\frac{\langle 0\vert(\bar{q}q)_\alpha\vert0\rangle\langle 0\vert(\bar{q}q)_\beta\vert0\rangle}{4F^4\chi_{\alpha\beta}
    (0)}\sum_{i,j}{\parent{\frac{c_i^\alpha\,
        c^\beta_j}{M_iM_j}\frac{\partial^2P}{\partial M_i\partial
        M_j}-\delta_{ij}\frac{c_j^{\alpha}c_j^{\beta}}{M_j^3}\frac{\partial P}{\partial M_j}}}\right].
\end{equation}

By considering the ratios $\langle(\bar{q}q)_\alpha\rangle_T/\langle(\bar{q}q)_\alpha\rangle_0$ and $\chi_{\alpha\beta}(T)/\chi_{\alpha\beta}(0)$, we cancel  the overall $B_0$ factors in the ChPT expressions for  susceptibilities and  condensates. These ratios are the quantities we will show in our plots. Note that, as it happened in the pure ChPT case, they are once again independent from the QCD renormalization group scale and can be expressed only in terms of meson parameters such as LEC, meson masses and decay constants. For these ratios
we still need
the values of $\langle 0\vert(\bar{q}q)_\alpha\vert 0\rangle/B_0$, for which we will take the one-loop ChPT expressions  from \cite{Gasser:1984gg}, and $\chi_{\alpha\beta}(0)/B_0^2$,  which
can be easily obtained from Eqs.~(\ref{suscsu3}) to (\ref{suscls})
taking $g_1=g_2=0$.

Recall also that the contribution to the susceptibility of the free part  of the pressure Eq.~(\ref{virial free pressure}) is precisely the same as the leading order ChPT results given in section \ref{sec:sus}, so that the size of the interaction contribution from Eqs.~(\ref{virial interaction coefficients})
and \eqref{Deltaphases} is a measure not only of the convergence of the virial series but also of the robustness of the pure ChPT contribution.

Finally, in order to evaluate
the interaction part of the virial coefficients, Eqs.~\eqref{virial interaction coefficients} and \eqref{Deltaphases}, we need the theoretical description of the meson-meson elastic scattering phase shifts,
which are nothing but the complex phase
of each scattering partial wave $t_{IJS}$. These partial waves are obtained
as the projection of the scattering amplitude in states of definite isospin $I$,
angular momentum $J$ and strangeness $S$. Let us remark that the unitarity of the $S$ matrix implies that, for physical values of CM energy squared $s$, partial
waves $t_{IJ}$  for \textit{elastic} meson-meson
scattering should satisfy:
\begin{equation}\label{unitarization}
  {\rm Im}\;t_{IJS}=\sigma
  |t_{IJS}|^2\;\;\;\Rightarrow\;\;\;{\rm Im}\frac{1}{t_{IJS}}=-\sigma\;\;\;\Rightarrow\;\;\;t_{IJS}=\frac{1}{\mathrm{Re}\,t_{IJS}^{-1}-\mathrm{i}\sigma},
\end{equation}
where $\sigma = 2p/\sqrt{s}$, and $p$ is the CM momenta of the two mesons. Note that
unitarity implies

\begin{equation}
t_{IJS}=\frac{\sin\delta_{IJS}}{\sigma}e^{i\delta_{IJS}},
\label{tsindelta}
\end{equation}
where $\delta_{IJS}$ are the phase shifts needed in Eq.~\eqref{Deltaphases}. The above equation leads to  $|t_{IJ} | \leq 1/\sigma$, and a strong interaction is characterized
precisely by the saturation of this unitarity bound.

In what follows we will first explain why the results for the susceptibility
obtained combining
standard ChPT with the virial expansion are
very uncertain at very low temperatures due to a huge cancellation between the interactions in the scalar channels, and
not reliable at moderate temperatures, since they do not describe data above typically 500 MeV. Later on we will explain
how this can be solved  using unitarized elastic ChPT, which provides a realistic description of data up to roughly 1 GeV.

\subsection{ChPT Phase Shifts}

Let us then start discussing the partial waves $t_{IJS}$, which
are obtained within standard ChPT as an expansion in even powers of momenta and meson masses. Dropping
for simplicity the $IJS$ indices, we find: $t(s)=t_2(s)+t_4(s)+\cdots$ where
$t_n(s)=O(p^n)$.
For our purposes it is enough to consider the one-loop $\pi\pi$ scattering calculation \cite{Gasser:1983yg}, i.e. up to $O(p^4)$, because, as it was already shown in \cite{Gerber:1988tt}, once it is included in the virial expansion and re-expanded in powers of temperature, it
reproduces the NNNLO pressure calculation, which is even more than the order considered in the previous sections. Hence, this is a simple procedure to estimate higher order thermal effects without performing the detailed $O(T^8)$ calculation for the susceptibilities.
Let us nevertheless recall that the ChPT series in
only valid at low energies compared  with $4\pi F \sim 1.2$ GeV, and in practice
it is limited to scattering momenta of the order of 200-300 MeV above threshold, or about 400-500 MeV in energy. The reason is that, experimentally, for larger momenta several partial waves become resonant, a behavior that cannot be reproduced with a power expansion in energy.

\begin{figure}
 \includegraphics[scale=.4]{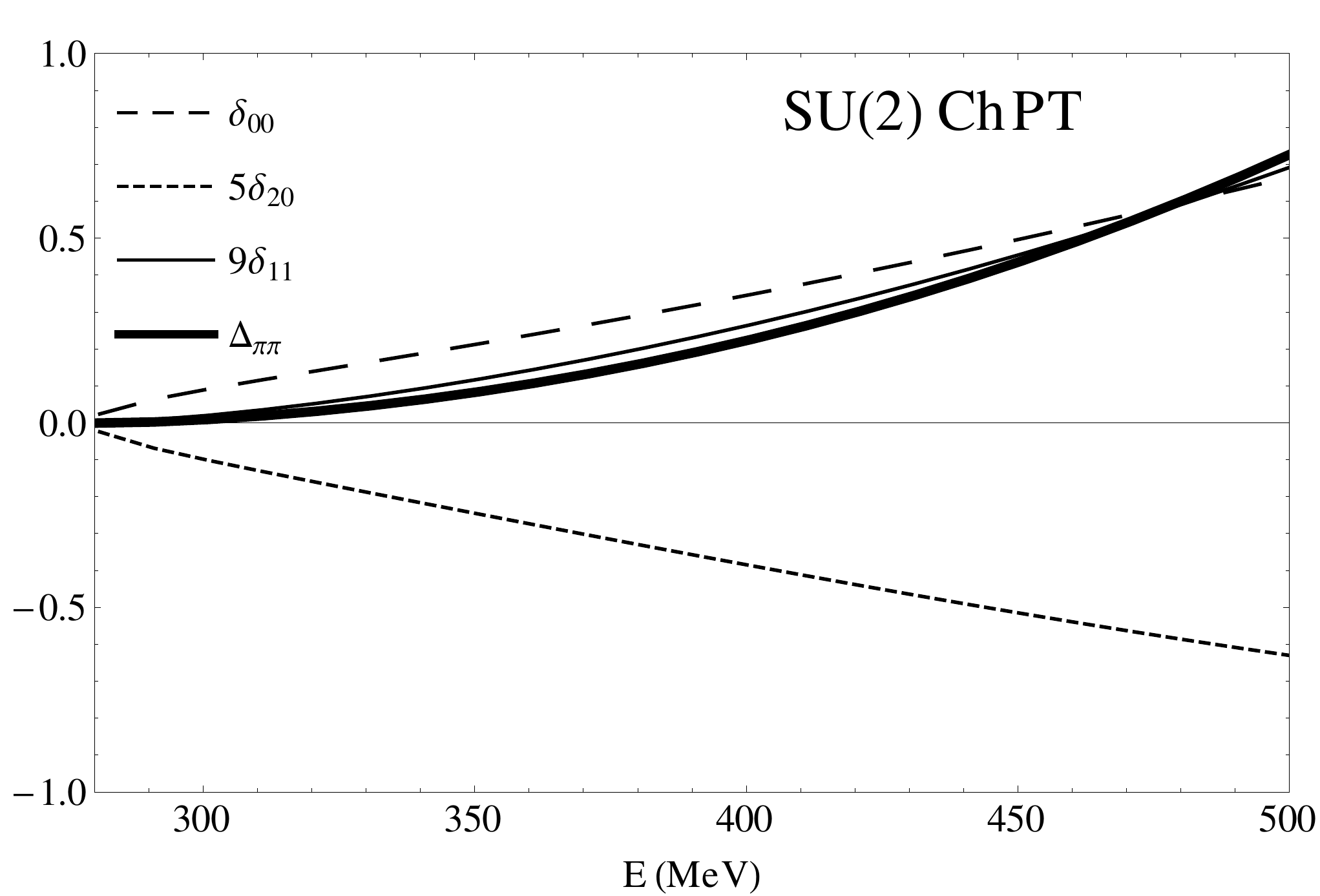}
 \includegraphics[scale=.4]{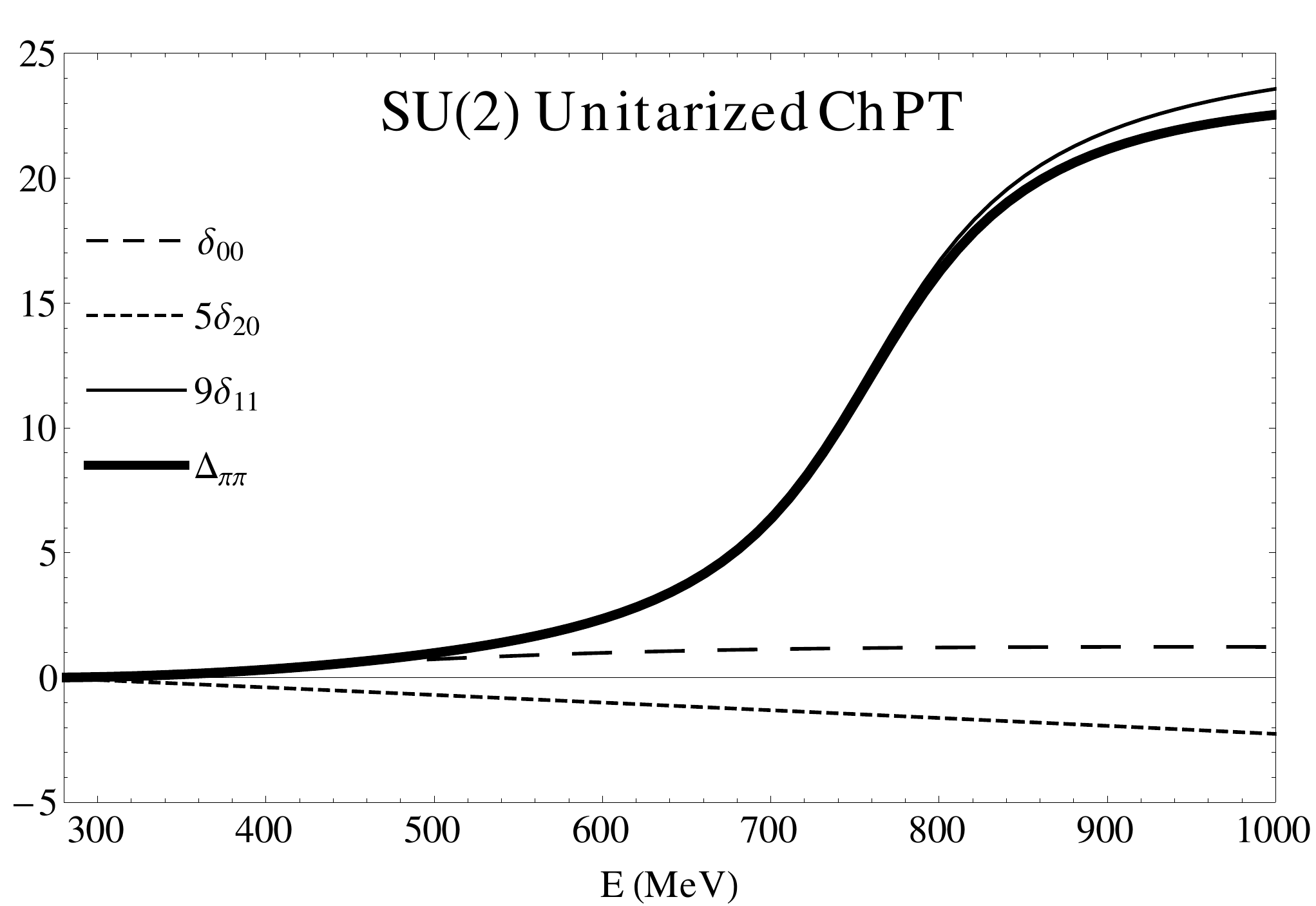}
 \includegraphics[scale=.4]{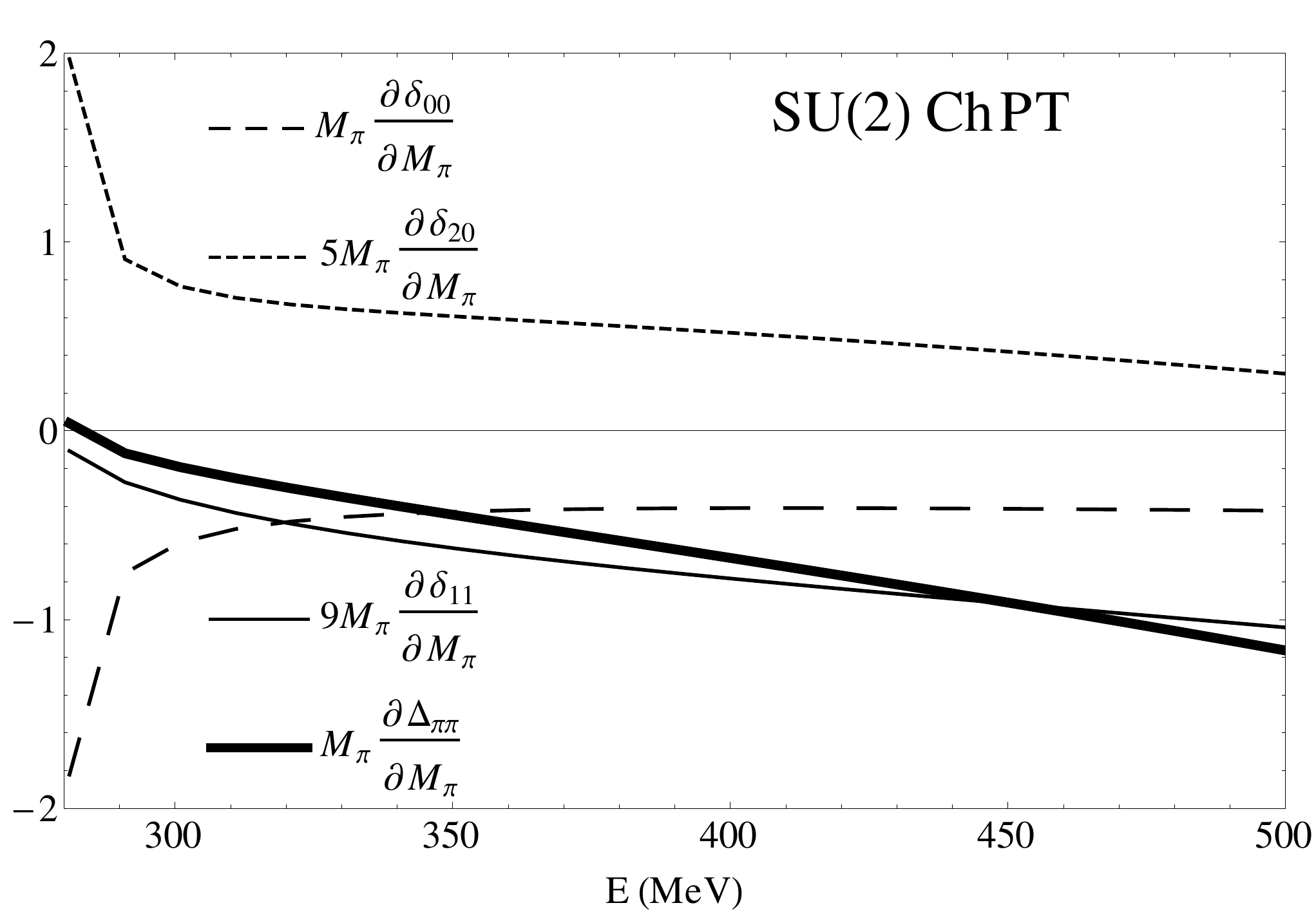}
 \includegraphics[scale=.4]{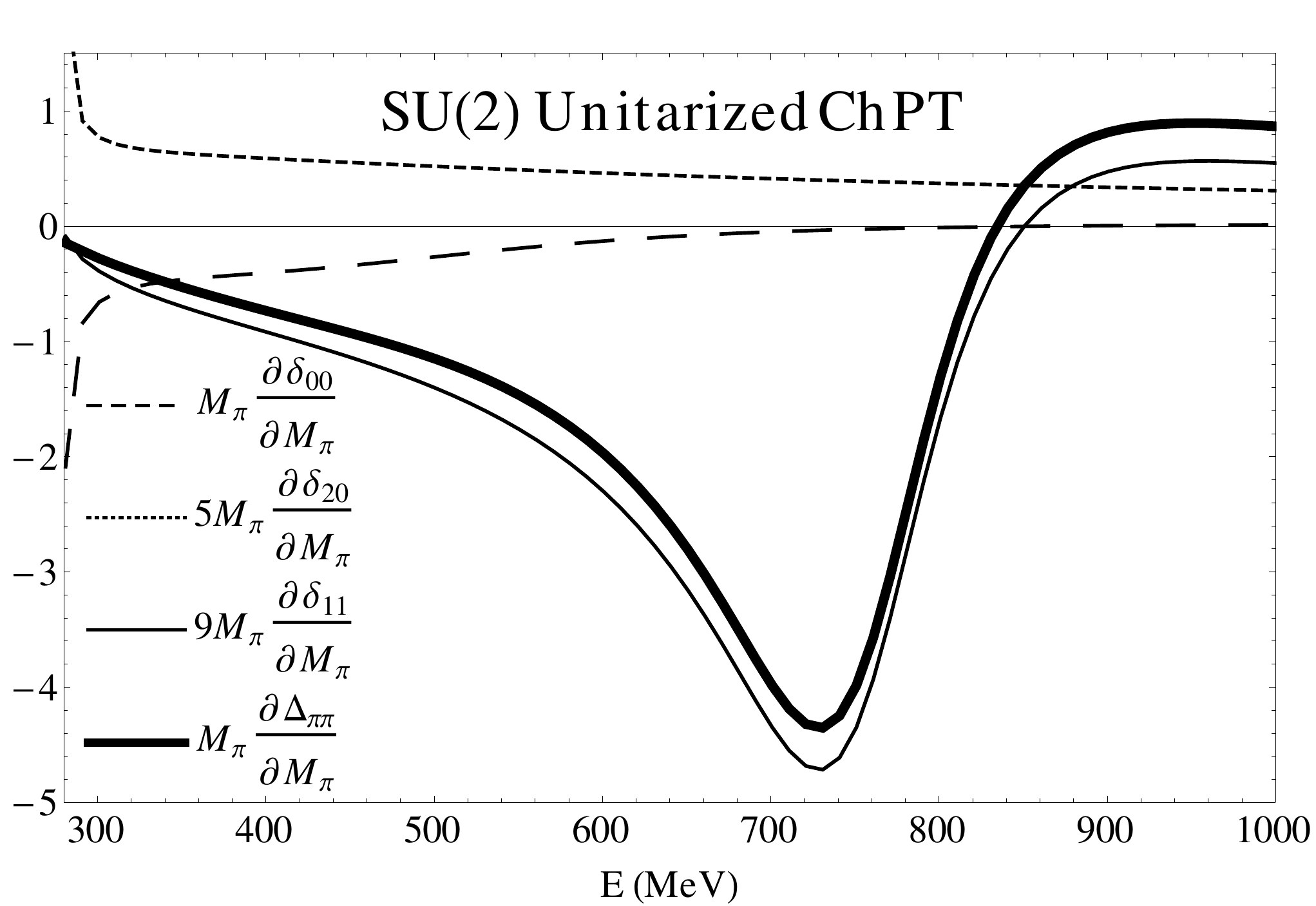}
 \includegraphics[scale=.4]{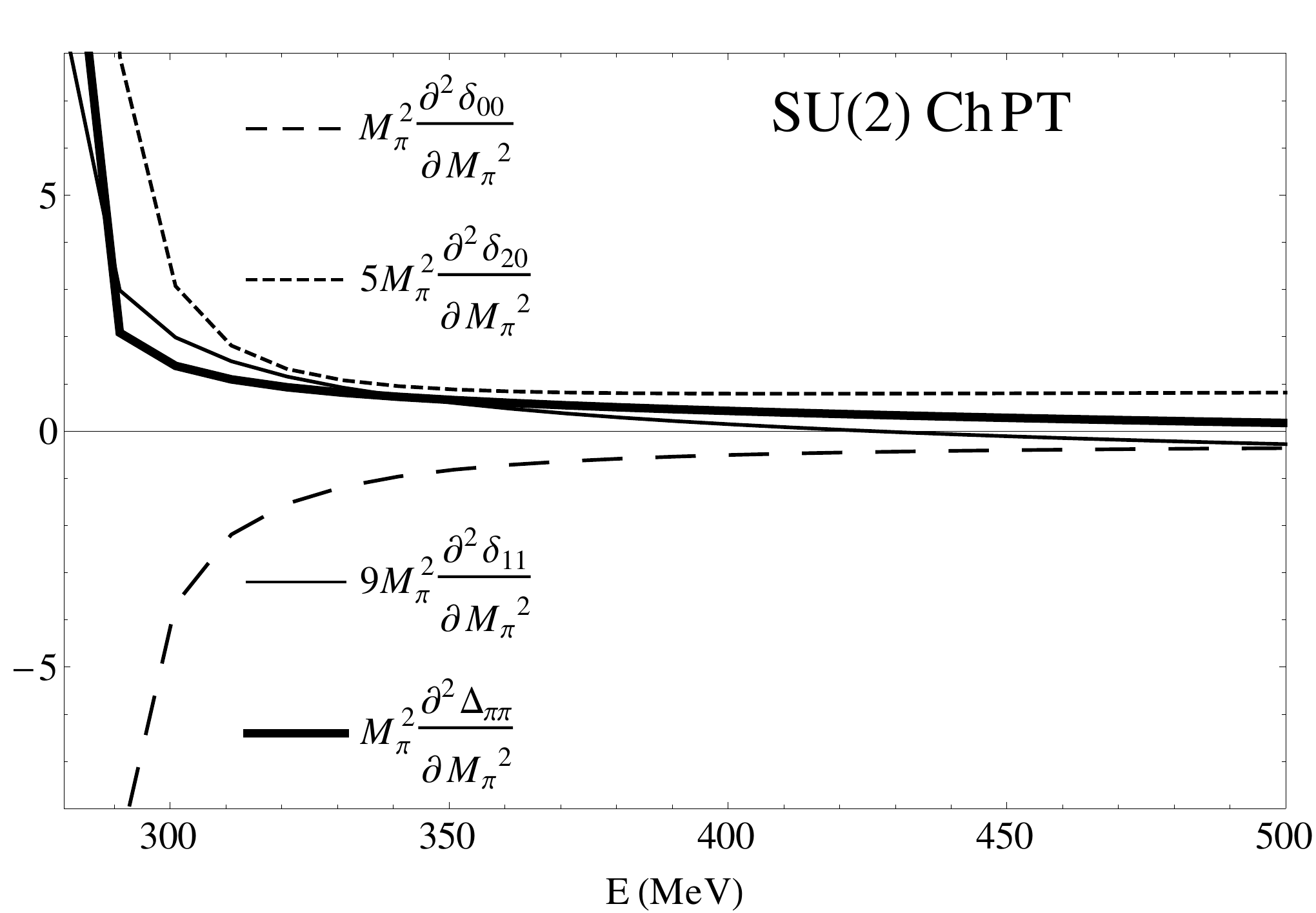}
 \includegraphics[scale=.4]{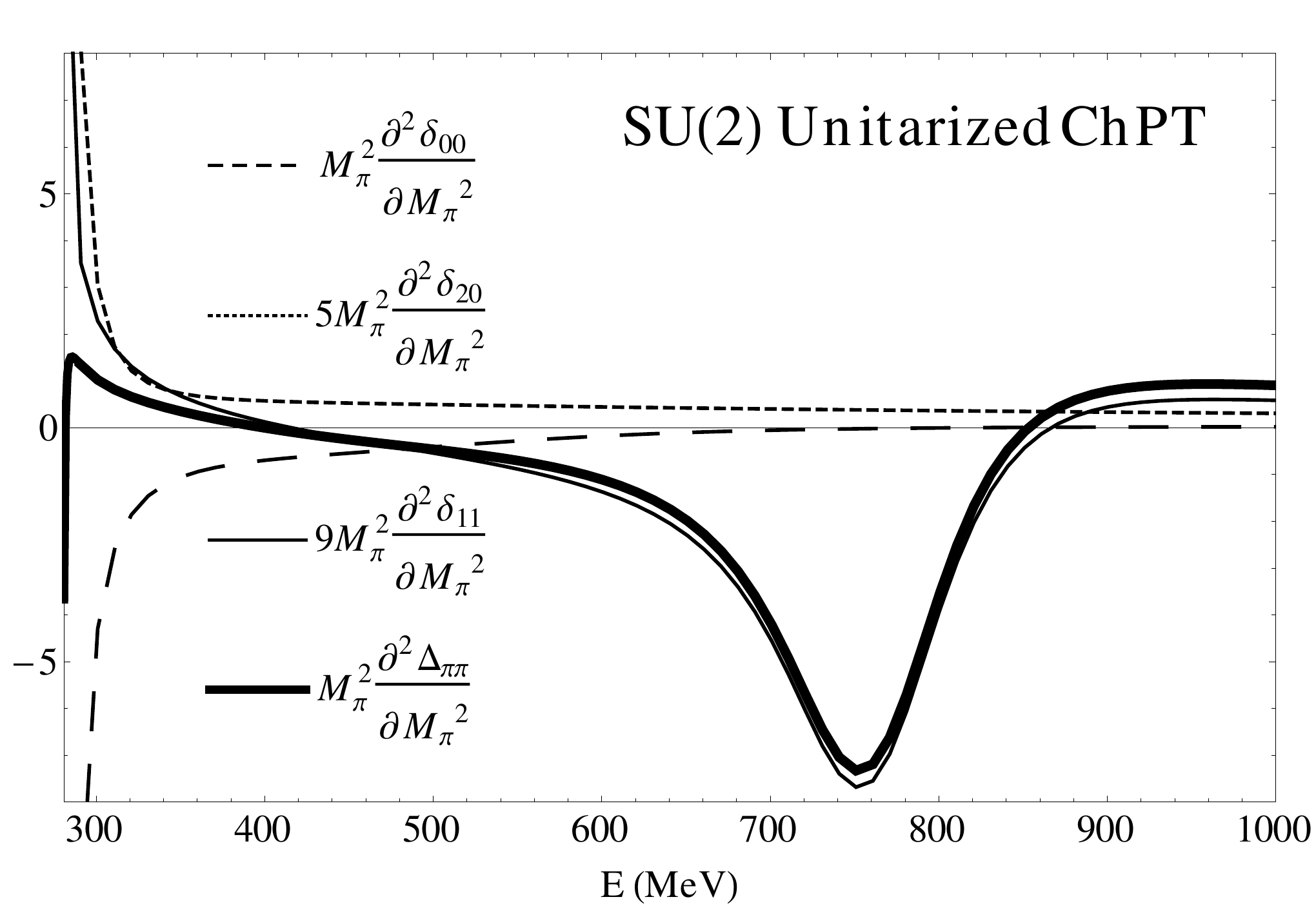}
  \caption{\rm \label{fig:cancellation}
From top to bottom, the $\pi\pi$
 scattering phase shifts $(2I+1)(2J+1)\delta_{IJ}$ and their first and second mass derivatives
compared to the combination $\Delta_{\pi\pi}=\delta_{00}+5\delta_{20}+9\delta_{11}$
and its mass derivatives. On the left column we plot the ChPT results and on the right one the unitarized ChPT calculations. Note the huge cancellation that occurs, in all graphs and irrespective of unitarization, between the $(I,J)=(0,0)$ and $(2,0)$ contributions.
Hence, $\Delta_{\pi\pi}$ and its derivatives are dominated by the $(1,1)$ contribution, with only a few exceptions around the threshold region where the $(1,1)$ channel suffers an additional suppression.
}
 \end{figure}

The energy integrals that define the virial coefficients, Eq.~\eqref{virial interaction coefficients}, extend to infinity, where the low energy ChPT expansion
is no longer valid. In principle, one may think that this is not a severe problem because, for the temperatures we are interested in, the very high energy region should be suppressed by the thermal Bessel functions in the virial coefficients.
However, as already explained, in the interaction
part of the virial coefficients,
Eq.\eqref{virial interaction coefficients}, the $\pi\pi$ scattering phase shifts appear
through the combination
\begin{equation}
\Delta_{\pi\pi}(E)=\delta_{00}(E)+5\delta_{20}(E)+9 \delta_{11}(E)+...
\end{equation}
where we have omitted waves with $J>2$ because they are significantly smaller than those with $J\leq1$ below 1 GeV, which is the region of interest for our calculations. It is now important to remark that, as we show in the top left panel of Fig.~\ref{fig:cancellation}
 there is huge cancellation between $\delta_{00}$ and $5\delta_{20}$. This cancellation was already observed for the scattering lengths in \cite{Gerber:1988tt} and in \cite{Venugopalan:1992hy} between attractive and repulsive channels. Using more recent determinations of the scattering lengths, we find
$a_{00}+5 a_{20}=0.002\pm0.009$ versus $a_{00}=0.220\pm0.005$ from \cite{Colangelo:2001df}, or  $a_{00}+5 a_{20}=0.010\pm0.015$ versus $a_{00}=0.220\pm0.008$ from \cite{GarciaMartin:2011cn}, namely, a cancellation of more than one, possibly two, orders of magnitude. Note that the last two numbers come from a data analysis that does not use ChPT.

As a consequence of that cancellation,  the resulting $\Delta_{\pi\pi}$ is basically given by $\delta_{11}$,
which at very low energies is extremely small. A similar cancellation also occurs both for the first and second mass derivatives of the phase shifts. Thus
$\partial \Delta_{\pi\pi}/\partial M_\pi$ and
$\partial^2 \Delta_{\pi\pi}/\partial M_\pi^2$
are strongly dominated by the contribution from the vector channel, as we show in the left middle and left bottom panels of Fig.\ref{fig:cancellation}. Let us nevertheless remark that the $J=1$ wave is suppressed at low energies by a $q^{2J}$ factor, and by itself yields a very small contribution to $\Delta_{\pi\pi}$ or its two first derivatives near threshold. Therefore, the model independent prediction of ChPT is that, at very low energies, the interaction part of the virial coefficient will be much smaller than naively expected from the size of each individual $J=0$ wave. In other words, the results of the virial approach with standard ChPT interactions at very low temperature follow the free gas approximation much closer than naively expected.

 Unfortunately, within ChPT we can state little more than the smallness
of the interaction part of the virial coefficient at very low temperatures,
because its value cannot be pinned down with numerical precision, since at very low energies what is left of $\Delta_{\pi\pi}$ after the cancellation is even smaller than the size of higher order corrections. Still, in the next section, we will try to estimate
those higher order contributions by means of unitarized ChPT.

Nevertheless, one could think about providing ChPT results for moderate temperatures, say 100 or 150 MeV, where the bulk of the contribution to the integral extends beyond the threshold region and reaches, say, 500 MeV. However, in view of the left column in Fig.\ref{fig:cancellation}, it is clear that this can be done for the pressure and its first derivative, i.e. for the condensate, because $\Delta_{\pi\pi}$ and $\partial \Delta_{\pi\pi}/\partial M_\pi$
are growing functions of the energy, and soon enough
the $\delta_{11}$ becomes dominant and the large systematic uncertainties near threshold become less relevant. But this is not the case for $\partial^2 \Delta_{\pi\pi}/\partial M_\pi^2$, which is needed for the susceptibility calculation. Actually, as seen in the left bottom panel of Fig.\ref{fig:cancellation}, this second mass derivative is a decreasing function of the energy, so that the region nearby threshold, with its huge systematic uncertainties, continues to be dominant even for moderate temperatures, within ChPT calculations.

Of course, if we consider even higher temperatures
we find the usual caveats for a standard ChPT calculation, but this time even more severe.
First of all because due to the cancellation commented above, the dominant contribution comes from the $\delta_{11}$ channel, which is also suppressed near threshold, and we can only find a sizable contribution when this $11$ channel becomes sufficiently large, but that only happens around the $\rho(770)$ resonance region. Unfortunately, whereas
the ChPT description of the $(I,J)=(0,0)$ channel
 is fairly good, at least qualitatively, up to energies as high as 700 MeV, the presence of the
$\rho(770)$ resonance in the $(I,J)=(1,1)$ channel is not reproduced even qualitatively.

Furthermore, the virial interaction coefficients in Eq.~(\ref{virial interaction coefficients}) are derived for an exactly unitary $S$ matrix \cite{Dashen,Welke:1990za}, and this high energy thermal suppression quickly takes place for physical amplitudes respecting unitarity. But at this point we recall that ChPT scattering partial waves
only satisfy unitarity perturbatively,
i.e: $\mathrm{Im}\,t_2 = 0$, $\mathrm{Im}\,t_4 = \sigma t_2^2$,... and
the unitarity constraint in Eq.\eqref{unitarization}
is badly violated pretty soon.
Moreover, this violation grows fast with increasing momenta or in the vicinity of resonances.
Hence, the contribution of ChPT
is unphysically large due to the unitarity violation of the perturbative expansion and
the contributions from the energy region $E>1\,$GeV, where any extrapolation is meaningless,
become sizable already at temperatures of 150 MeV, for the ChPT virial calculation.
In other words, for small temperatures,
say formally $T\ll M_\pi$ the relevant momenta in the virial integrals
are $p\sim\sqrt{M_\pi T}$ and hence the amplitudes are probed in
the range $p\ll M_\pi$ where ChPT can be trusted. However, for $T\gsim M_\pi$
momenta are of the order of $2M_\pi$ or greater and the extrapolation of standard
 ChPT amplitudes does not describe meson scattering data.

Within one-loop ChPT, the correct low energy expansion of the non-unitarized phase shift
is $\delta_{NU}\simeq\sigma (t_2+\mbox{Re}t_4)$, and gives a reasonable description of experimental data for very low energies with our choice of LEC. Then,
for $E\gg M_\pi$, $\delta_{NU}(E)\sim E^4$, $\partial \delta_{NU}(E)/\partial M_\pi\sim E^2$, which produces additional powers of $T$ in the interaction part compared to the free contribution, giving a large but unphysical weight to the
higher energy contributions to the virial coefficients. Note in turn that, by this same argument, second derivative terms are subdominant in Eq.~(\ref{chivirial}) with respect to the first derivatives. A rough estimation of the asymptotic behavior with $T$, say formally $T\gg M_\pi$, can be obtained by looking at the $E\gg M_\pi$ behavior of the integrand in Eq.~(\ref{virial interaction coefficients}). The thermal function $x^2 K_1(x)$ weights the region $x\sim 1$ so that asymptotically we can just trade factors of $E$ in the phase shifts by $T$, which also allows to compare the interaction part in Eq.~(\ref{virial interaction coefficients}) with the free contributions Eq.~(\ref{virial free coefficients}).

For all of these reasons, next we will make use of the so-called unitarized ChPT, which, under some reasonable approximations, combines ChPT at low energies with dispersion relations, and provides a realistic description of the data, reproducing the resonances relevant for this work, without introducing any spurious parameter other than those of ChPT. Of course, the price to pay is the loss of the systematic ChPT approach.

\subsection{Unitarized interactions}
\label{sec:unitin}

In order to check the influence in the virial integrals
of higher ChPT orders, the violation of unitarity and
the lack of resonances, we will extend the ChPT amplitudes by means of unitarization, and in particular, we will use the elastic Inverse Amplitude Method (IAM) \cite{Truong:1988zp,Dobado:1992ha} which provides a remarkably
good description of the meson-meson scattering data up to roughly 1 GeV.

Unitarization methods provide amplitudes up to higher energies by using
the fact implicit in Eq.~\eqref{unitarization}, that \textit{the imaginary part of the inverse amplitude
is known exactly}. Naively, we can then impose the ChPT constraints to the real part of ${\rm Re}\,t^{-1}
\simeq t^{-2}_2 (t_2 + {\rm Re}\,t_4 + ...)$ to find that
\begin{equation}\label{t IAM}
t = \frac{1}{{\rm Re}t^{-1}- i\sigma}\simeq\frac{t_2}{1-t_4/t_2}.
\end{equation}

This is the one-channel IAM \cite{Truong:1988zp,Dobado:1992ha,GomezNicola:2007qj}. Although the use of the ChPT series in this
naive derivation is only valid at low energies, the IAM can be
derived also from a subtracted dispersion relation for the inverse amplitude,
which justifies its extension to higher energies and even to the complex plane.
The details of the dispersive derivation can be found in
\cite{Dobado:1992ha,GomezNicola:2007qj}, but for our purposes here it is important to remark that the elastic cut is calculated exactly thanks to unitarity and the subtraction constants are calculated with ChPT, which is well justified since they correspond to evaluating the amplitude at very low energies. The IAM equation is valid at any energy as long as the left cut integral is well approximated  by its low energy ChPT expansion, which is justified due to the subtractions, and as long as the energy where the amplitude is evaluated is sufficiently far from the inelastic region. Other terms due to so-called Adler zeros have also been shown explicitly to be negligible \cite{GomezNicola:2007qj} in this region.

In summary, following a more rigorous derivation than the naive one above, the very same Eq.~\eqref{t IAM} is recovered not only at low energies, but for most of the elastic region. Remarkably, this simple equation is able to describe
 meson-meson scattering data, enlarging considerably the energy applicability range \cite{Dobado:1992ha}, while still reproducing the ChPT series at low energies.
In addition, the IAM generates \cite{Dobado:1992ha} the poles in the second Riemann sheet associated with the resonances, from first principles like unitarity, analyticity and the QCD chiral symmetry
breaking, without introducing these resonances by hand or any spurious parameter beyond the LEC of ChPT. Thus, the low-lying meson resonances are well described with this method, in very good agreement with the existing data.  All these features can be reproduced  with values
of the ChPT parameters (LEC) that are fairly compatible with the values obtained within
standard ChPT, despite being obtained from a fit to a much larger energy region. Nevertheless is important to point out that due to the nature of the IAM approach, the LEC needed to fit data and resonance poles with these unitarized amplitudes are only approximately those of ChPT. For this reason, we will use for the IAM phase shifts in the virial expansion the set of LEC obtained by fitting both scattering data and lattice results on meson masses, decay constants and scattering lengths. For the SU(3) case
we take the values in \cite{Nebreda:2010wv} and for the SU(2) case from \cite{Nebreda:2011di}.
Note that we will only unitarize the pion-pion scattering amplitude since, as explained in previous sections and as it can be seen in our figures, this is the dominant contribution from meson interactions and, as we will see, correcting it with the IAM gives a considerably larger effect than including or not the kaon and eta interactions. Actually, and for the sake of simplicity, we will only consider two unitarized situations: either $SU(2)$ or $SU(3)$ but considering only free kaons and etas.
With a lesser degree of rigor, the IAM can even be extended to the inelastic region \cite{GomezNicola:2001as} above 1 GeV, although that regime is not relevant for this work and the elastic formalism is enough for our purposes. Thus we prefer to rely on the most rigorous elastic formalism obtained from dispersion theory.

Thus, in the right column of Fig.~\ref{fig:cancellation} we show the $\pi\pi$ phase shifts and their derivatives which are obtained from our unitarized ChPT (UChPT) calculations. Once again we find a huge cancellation between the $(I,J)=(0,0)$ and
the $(2,0)$ contributions.  In the case of the phase and its first derivative, which have less uncertainty in the cancellation and also become very small at threshold, the effect of this cancellation uncertainty is very small in the virial integrals. But this is not the
case for $\partial^2\Delta_{\pi\pi}/\partial M_\pi^2$,  as can be noticed when
comparing the lower left and right panels, where we see that there are large uncertainties near threshold due to the higher order effects. Note that the change on each individual wave due to unitarization is rather small, but a mere 10\% change in the $(0,0)$ channel produces a change of sign in the unitarized $\partial^2\Delta_{\pi\pi}/\partial M_\pi^2$. Hence, the interaction contribution to the susceptibilities is rather uncertain, but the overall uncertainty at very low energies is still small
since the free contribution dominates by large.

Note also that, for the unitarized case, we now draw the phases up to 1 GeV in order to show the almost complete dominance of the $\rho(770)$ resonance contribution to $\Delta_{\pi\pi}$ and its derivatives above $E=500\,$MeV. The previously commented cancellation between the $(0,0)$ and $(2,0)$ contributions still persists at low energies, but deteriorates slightly above 500 MeV, where the $\rho(770)$ contribution simply dominates because it is much larger than the others. This $\rho(770)$ dominance is very relevant to asses the reliability of the UChPT results, since it has been recently shown that the $\rho(770)$ mass dependence obtained with the one-loop IAM is in fairly good agreement with the most recent lattice calculations \cite{qmassUChPT}.

If we now recall the result that in the narrow width approximation a resonance exchange contributes to the partition function as the free resonance state would do \cite{Dashen}, we conclude that the usual HRG with a free $\rho(770)$, is, according to our results, a fairly consistent approach to include the $\pi\pi$ interactions. In contrast, one might naively expect that the $\sigma$ resonance, which is the nearest one to threshold and also has the quantum numbers of the vacuum, should provide the largest contribution to the susceptibility. However,
and this  is one of the remarkable results of this work, we have shown that in the threshold region it suffers a dramatic cancellation with the $(I,J)$=$(2,0)$ interaction. Therefore including just the $\sigma$ as a free state in a HRG without the $(2,0)$ interaction, apart from ignoring the fact that the $\sigma$ is by no means a narrow state, also neglects this very important cancellation.

Moreover, the unitarized partial waves  have a much softer behavior for large energies, namely  $t(E)$ behaves as a constant, giving rise to the asymptotic behavior $\delta_U (E)\sim$ constant, $\partial \delta_{U}(E)/\partial M_\pi\sim 1/E^2$. Thus, for the susceptibility in Eq.~(\ref{chivirial}), the interaction part is suppressed with respect to the free one by inverse powers of $T$ and the result is driven by the ChPT one. We have explicitly checked that the contributions to the integrals from energies higher than 1 GeV are very suppressed now, and barely affect our results, contrary to the non-unitarized case.

\begin{figure}
 \includegraphics[scale=.5]{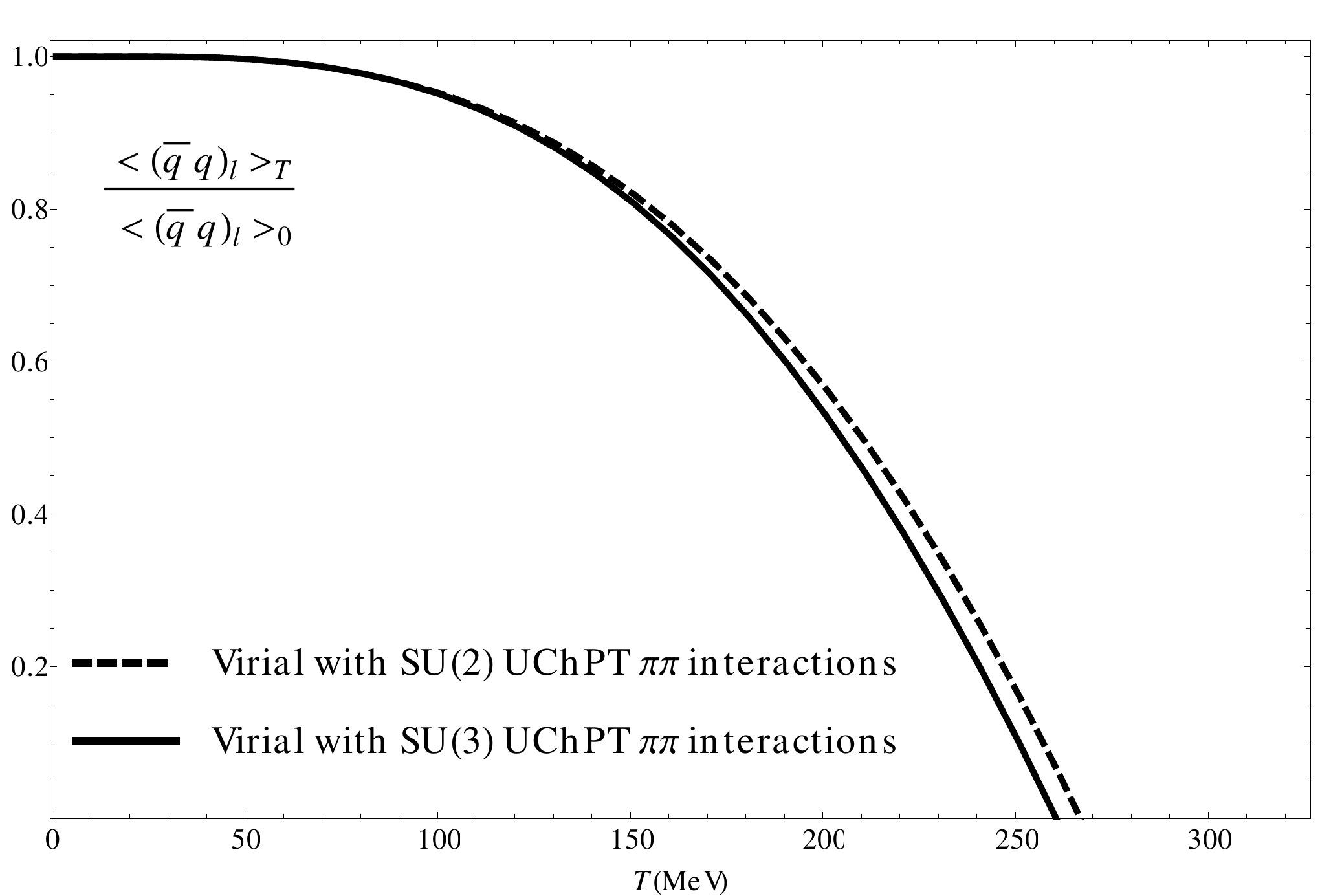}
  \caption{\rm \label{fig:condvirU} Non-strange quark condensate as a function of temperature in the virial approach, using Unitarized interactions in SU(2) (dotted line)and in SU(3), only with pion-pion interaction and free kaons and etas (solid line)}
 \end{figure}

Thus, in Fig.~\ref{fig:condvirU} we plot the non-strange quark condensate using the virial approach with unitarized $\pi\pi$ interactions both within the SU(2) and SU(3) formalisms. The extrapolated melting temperatures are somewhat lower than those coming from standard NNLO ChPT calculations, already given in  Fig.~\ref{fig:condchpt}, particularly for the SU(2) case, this is partly explained, since as we have just discussed we are adding, in practice, the $\rho(770)$ as an additional degree of freedom. Note also that the paramagnetic decrease between the SU(2) and SU(3) cases is just of the order of 6 MeV, which is smaller than the one estimated with NNLO ChPT. Nevertheless this smaller difference is less reliable since it is not calculated with SU(2) LEC obtained from those of SU(3), since the unitarized phases are obtained by fitting to different sets of data in both cases.

Taking into account all the above considerations we have also plotted in Fig.~\ref{fig:suscChPT} the chiral susceptibility with unitarized pion interactions. The results  are similar to NNLO ChPT,
and the difference provides the crude estimate of systematic uncertainties, giving
 rise to a quite consistent picture between ChPT and the virial approach.

\section{Comparison with lattice}
\label{sec:latcom}

As explained in the introduction, the light scalar susceptibility
is one of the main parameters studied by many lattice collaborations \cite{Bernard:2004je,Aoki:2006br,Aoki:2006we,Cheng:2009zi,Aoki:2009sc,Ejiri:2009ac,Borsanyi:2010bp} in order to identify its peak position as the transition point . A suitable quantity we can compare with is
 $\Delta(T)\equiv m^2\left[\chi_l(T)-\chi_l(0)\right]/M_\pi^4=\left[\chi_l(T)-\chi_l(0)\right]/(4B_0^2)$,
which is given  for instance in \cite{Aoki:2009sc} for  2+1 $SU(3)$ flavor simulations with almost physical quark masses. By subtracting the $T=0$ value, the lattice analysis of this quantity is free of ultraviolet divergences.  Besides,  $\Delta(T)$ obtained from the perturbative ChPT
result in Eq.~(\ref{susclightsu3}) is not only independent of $B_0$ but also of  the LEC (in particular of $H_2$, which is subject to more uncertainty, as explained above). Thus, the ChPT result for $\Delta(T)$ depends only on meson masses and  temperature.  In the virial case, Eq.~(\ref{chivirial}), there is no $B_0$ dependence but the result still depends on the LEC, through the $T=0$ condensates, masses and phase shifts.

\begin{figure}
\includegraphics[scale=.6]{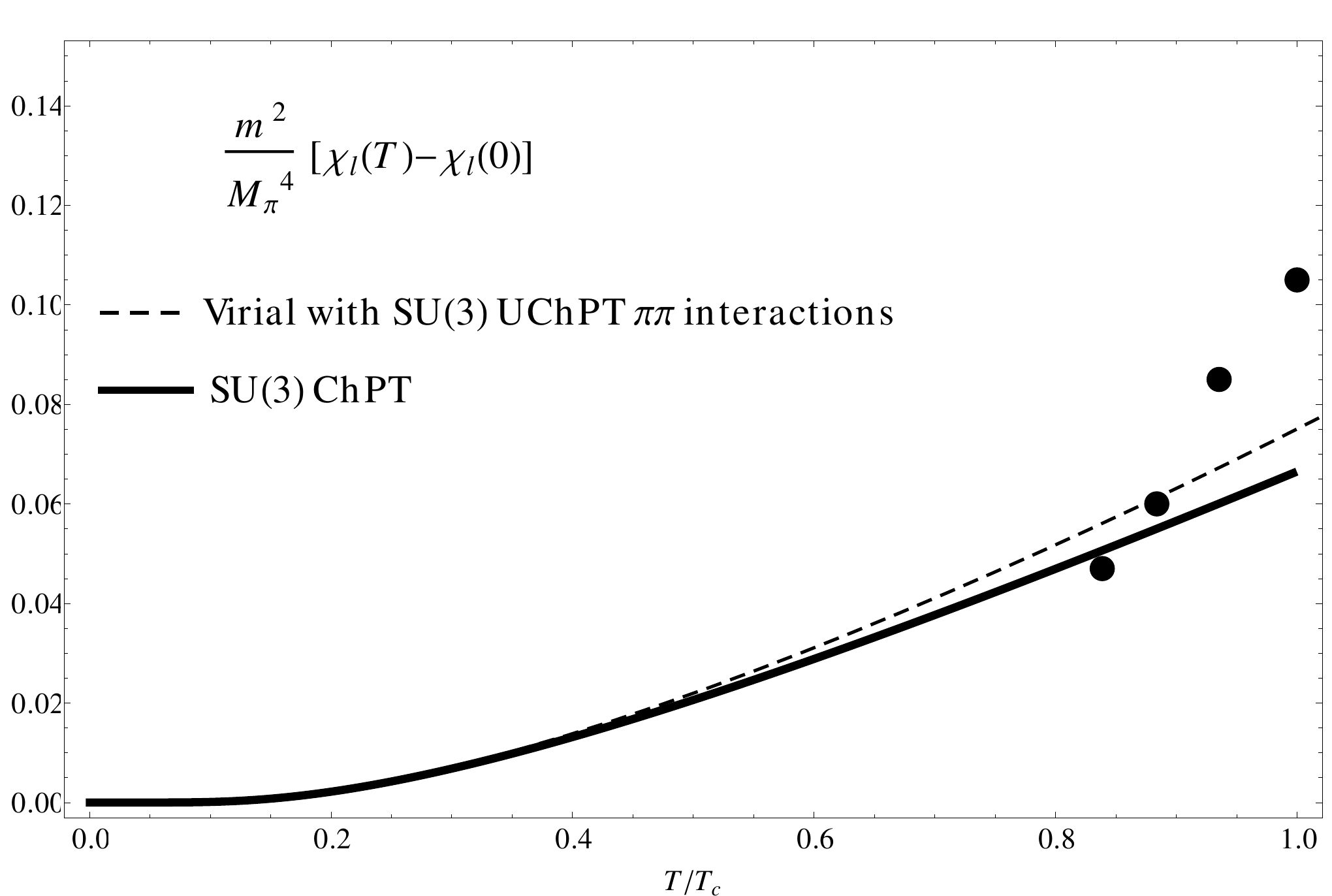}
 \caption{\rm \label{fig:susrel} Normalized relative non-strange scalar susceptibility in $SU(3)$ in terms of the relative temperature, in perturbative ChPT  and in the unitarized virial approach. The points are lattice data taken from \cite{Aoki:2009sc}, where $T_c\simeq$ 155 MeV.}
\end{figure}

As discussed previously, the lattice results predict a critical temperature considerably smaller than ChPT or virial extrapolations. This corresponds in part to the relevance of degrees of freedom of higher masses near $T_c$. Qualitatively, one expects that those degrees of freedom produce a ``paramagnetic" reduction of $T_c$ due to the increase of entropy. Thus, in order to establish a more appropriate comparison with lattice results, we will represent the results in terms of the reduced temperature $T/T_c$ for each approach (ChPT, virial and lattice) which is a way to compensate for the number of degrees of freedom involved.

The results for $\Delta(T)$ are plotted in Fig.~\ref{fig:susrel}. At very low $T$,  the  curves remain close to one another. At high temperatures they grow roughly linear in $T$,
up to the critical region.
As stated before, this is a check of the robustness of both approaches, since the unitarized virial result is not only obtained within a different framework but it includes dependencies on the unitarization method and the LEC, thus giving an estimate of systematic uncertainties.

Compared to the lattice data, we see that the two lowest points available are rather well
described with the ChPT or unitarized virial curves. This is reassuring, since ChPT is meant to provide the low $T$ model-independent tail. Once the temperature is re-scaled to $T_c$ we see that the agreement is rather good even up to $0.9 T_c$, which is remarkable, given that ChPT captures only the qualitative features of the evolution towards chiral restoration but does not develop for instance a maximum, not even a sudden increase of $\Delta(T)$ near $T_c$. Actually, the  lattice points reflect a clear departure from the ChPT prediction as they approach  the critical point.

\section{Conclusions}

We have analyzed several properties of the meson gas at low temperatures, regarding four-quark condensates and susceptibilities, within the ChPT and virial expansion approaches. Our analysis provides helpful results for understanding  the behavior of the hadron gas formed after a relativistic heavy ion collision, below the chiral phase transition.

The factorization hypothesis for four-quark correlators does not hold in ChPT at finite temperature to NNLO in the chiral expansion. This is an extension of a previous $T=0$ analysis and is a model independent result. In particular, it means that, in the physical case,
the four-quark scalar condensate cannot be used as an order parameter for chiral restoration, since it contains divergent factorization breaking terms that cannot be renormalized.
Nevertheless, there are two particular limits in which the four-quark condensate could be considered an order parameter, namely, the the large $N_c$ limit, where
factorization holds formally, and the chiral limit, where factorization is still broken, but, contrary to the $T=0$ case, just by a finite contribution. Let us remark that the factorization breaking terms are precisely those needed to provide finite and scale-independent scalar susceptibilities for the light and strange sector, including the mixed one. We have provided explicit expressions for those susceptibilities in ChPT to leading order in their chiral expansion. The most important one, regarding chiral restoration, is the light scalar susceptibility, which grows linearly in $T$ at low temperatures.

In order to establish properly the previous factorization results, we have calculated the two-quark condensate in $SU(3)$ up to NNLO in ChPT, including meson interactions with strange  degrees of freedom ($\pi K$, $\pi\eta$, $K\eta$ and so on). In particular,  this allowed us to discuss the effect of those interactions in the
crude determination of $T_c$ from the extrapolated condensate. In addition, since the LEC of fourth order enter at that level, we have been able also to estimate the influence of the LEC in the ChPT determination of $T_c$. The net effect of strange interactions is about $\Delta T_c = 28\pm 12$ MeV where the error comes from the LEC uncertainty.

An important part of our work has been devoted to the comparison of the ChPT approach with the virial or density expansion, especially for the light scalar susceptibility. We have shown that one should consider unitarized interactions when probing the low and moderate temperatures of interest for this work. Moreover, the unitarized virial curves remain close to the standard ChPT calculation, which is a reflection of the robustness of both approaches, at least at low and moderate $T$,  and shows that the pion interactions are suppressed in thermal observables. This suppression is even larger than expected due to a huge cancellation between the scalar channels with isospin 0 and 2, which is already observed in the data below 1 GeV, but we have found also to occur for the first and second mass derivatives of the interactions. The existence of this cancellation implies that if the sigma particle, which dominates the scalar isoscalar channel near threshold is included alone as a free state, as in the usual hadron resonance gas approach, it can give large deviations from the model independent ChPT approach even at very low temperatures. As a consequence of this cancellation, the $\rho(770)$ dominates the interaction contribution starting at moderate temperatures of the order of 100 to 150 MeV, accelerating the melting of the condensate and the susceptibility growth.

The comparison with lattice data shows a remarkable agreement at those temperatures, when they are re-scaled with respect to the corresponding critical temperatures, i.e., compensating by the number of degrees of freedom involved in the calculation.   When we compare between the $SU(2)$ and $SU(3)$ cases, we do not see a significant difference between considering free or interacting kaons and eta compared to the effect of unitarizing the interactions. In the standard virial treatment, where the phase shifts are considered at $T=0$ and the thermal correction comes from the weight density functions, we have showed that the unitarized interactions decrease with inverse powers of $T$ with respect to the free contribution.

\appendix

\section{SU(2) results}\label{sec:su2res}
Here we collect, for completeness, the $SU(2)$ results for the four-quark correlator, factorization and scalar susceptibility.

For the four-quark correlator we get:
\begin{eqnarray}
{\quarkcor}_{T,\;NLO}\!\!&=&\condtwo_{T,\;NLO}^2,
\label{Qsu2NLO}\\
\quarkcor_{T,\;NNLO}\!\!&=&\condtwo_{T,\;NNLO}^2+B_0^2\left[-8 i(l_3+h_1)\delta^{(D)}(x)+K^T_{SU2}(x)\right]\label{Qsu2NNLO},
\end{eqnarray}
where we have defined $K_{SU2}^{T}(x)$ as the connected part of the four-pion correlator at leading order and finite temperature:
\begin{equation}\label{Ksu2}
K^T_{SU2}(x)=\langle {\cal T}\phi^a(x)\phi_a(x)\phi^b(0)\phi_b(0)\rangle_{T,LO}-\langle {\cal T}\phi^a(0)\phi_a(0)\rangle^2_{T,LO}=6G^T_\pi(x)^2.
\end{equation}
As  it happened in  $SU(3)$, to NLO the four-quark correlator can be expressed again as the square of the quark condensate, but not to NNLO where  $K^T_{SU2}$ breaks such factorization. As we did in the main text, we are giving  Eqs.~\eqref{Qsu2NLO} and \eqref{Qsu2NNLO} simplified in terms of the explicit expression for $\condtwo_{T,\;NNLO}$, which are given in Appendix~\ref{app:renquark}.

As for factorization, we have from Eq.~\eqref{Qsu2NNLO}:
\begin{equation}\label{factbreakingsu2}
\frac{{\condfourT}}{{\condtwoT}^2}=1+\frac{3}{2F^4}\parent{G_\pi(0)+g_1(M_\pi,T)}^2,
\end{equation}
which is again divergent with the standard ChPT renormalization \cite{GomezNicola:2010tb}.

The $SU(2)$ susceptibility is given by:
\begin{eqnarray}\label{su2sus}
\chi_l^{SU(2)}(T)&=&B_0^2\left[8\left(l_3^r(\mu)+h_1^r(\mu)\right)-12\nu_\pi+6g_2(M_\pi,T)\right]+\Od\left(\frac{1}{F^2}\right).
\end{eqnarray}

An important comment is that the renormalized $SU(2)$ LEC can be written in terms of the $SU(3)$ ones by performing formally an expansion for large strange quark mass in a given observable calculated in $SU(3)$ and comparing with the corresponding $SU(2)$ expression \cite{Gasser:1984gg}, although the numerical difference between them is small. In the $SU(2)$ case, the only two combinations of LEC appearing in the NNLO expression for $\condtwo_T/\condtwo_0$ are $l_3^r$ and $l_3^r+ h_1^r$, which can be readily expressed in terms of the $SU(3)$ LEC. Applying such LEC conversion in Eq.~(\ref{su2sus}) one gets directly the $SU(3)$ expression in Eq.~(\ref{susclightsu3})  if the $g_2(M_{k,\eta},T)$ are neglected.

\section{Finite temperature quark condensates to NNLO in ChPT}
\label{app:renquark}

In this appendix we will provide the NNLO results for the two-quark condensates at finite temperature.
As explained in the main text, the corresponding four-quark condensates
cannot  be obtained just by squaring these results, but one also has to add the factorization breaking contributions
described in Eqs.~\eqref{Qsu2NNLO} and \eqref{Qsu3NNLO}. The renormalization needed to render the quark condensates finite and scale independent is the same as for $T=0$. Therefore,  for all the technical aspects concerning the conventions for the needed ${\cal L}_4$ and ${\cal L}_6$ LEC and their renormalization, we refer to \cite{GomezNicola:2010tb}.

For convenience  we  define:

\begin{eqnarray}
\mu_i(T)=\frac{M_{0i}^2}{32\pi^2 F^2}\log\frac{M_{0i}^2}{\mu^2}+\frac{g_1(M_i,T)}{2F^2}, \qquad
\nu_i(T)=F^2\frac{\partial\mu_{i}(T)}{\partial M_{0i}^2}=\frac{1}{32\pi^2}\left(1+\log\frac{M_{0i}^2}{\mu^2}\right)-\frac{g_2(M_i,T)}{2},
\label{nudefpi}
\end{eqnarray}
where we denote, following the notation in \cite{Gasser:1984gg}, $\mu_i\equiv \mu_i(0)=\mu_i(g_1=0)$.

The final expressions for the  two-quark condensates, finite and
scale-independent, up to NNLO, at $T\neq0$ are given by:
\begin{eqnarray}
\condtwol^{SU(2)}_{NLO} (T)&=& -2B_0F^2\left\{1+\frac{2 M_{0\pi}^2}{F^2}\left(h_1^r+l_3^r\right) -3\mu_\pi(T)  \right\}, \label{condtwosu2nlo}\\
\condtwol^{SU(2)}_{NNLO} (T)&=&\condtwo^{SU(2)}_{l,NLO}-2B_0F^2\left[
-\frac{3}{2}\mu_\pi^2(T) -\frac{3 M_{0\pi}^2}{F^2}\left(\mu_\pi(T)\nu_\pi(T)+4l_3^r\mu_\pi(T)\right)\right.\nonumber\\
&+&\left.\frac{3M_{0\pi}^4}{8F^4}\left(-16l^r_3\nu_\pi(T)+\hat c_1^r\right)\right],
\label{condtwosu2nnlo}
\end{eqnarray}

\begin{eqnarray}
\condtwol^{SU(3)}_{NLO} (T)&=& -2B_0F^2\left\{1+\frac{4}{F^2}\left[\left(H_2^r + 4 L_6^r + 2 L_8^r\right) M_{0\pi}^2+8 L_6^r M_{0K}^2  \right]-3\mu_\pi(T) -2\mu_K(T)-\frac{1}{3}\mu_\eta(T) \right\},\label{condtwosu3nlo}\\
\condtwol^{SU(3)}_{NNLO} (T)&=&\condtwo^{SU(3)}_{l,NLO}-2B_0F^2\left\{
-\frac{3}{2}\mu_\pi^2(T) +\frac{1}{18}\mu_\eta^2(T)+\mu_\pi(T)\mu_\eta(T)-\frac{4}{3}\mu_K(T)\mu_\eta(T)\right.\nonumber\\
&+&\left.\frac{1}{F^2}\left[-3M_{0\pi}^2\mu_\pi(T)\nu_\pi(T)
+\frac{1}{3}M_{0\pi}^2\mu_\pi(T)\nu_\eta(T)-\frac{8}{9}M_{0K}^2\mu_K(T)\nu_\eta(T)\right.\right.\nonumber\\
&+&\left.\left. M_{0\pi}^2\mu_\eta(T)\nu_\pi(T)-\frac{4}{3}M_{0K}^2\mu_\eta(T)\nu_K(T)+\frac{1}{27}\left(16M_{0K}^2-7M_{0\pi}^2\right)\mu_\eta(T)\nu_\eta(T)\right]\right.\nonumber\\
&+&\frac{24}{F^2}\mu_\pi(T)\left[\left(3 L_4^r + 2 L_5^r - 6 L_6^r - 4 L_8^r\right)M_{0\pi}^2+2\left(L_4^r - 2 L_6^r\right)M_{0K}^2\right]
\nonumber\\&+&\frac{16}{F^2}\mu_K(T)\left[\left(L_4^r - 2 L_6^r\right)M_{0\pi}^2+2\left(3 L_4^r +  L_5^r - 6 L_6^r - 2 L_8^r\right)M_{0K}^2\right]\nonumber\\
&+&\frac{8}{9F^2}\mu_\eta(T)\left[\left(-3 L_4^r - 2 L_5^r + 6 L_6^r - 48 L_7^r - 12 L_8^r\right)M_{0\pi}^2+2\left(15 L_4^r + 4 L_5^r - 30 L_6^r + 24 L_7^r\right)M_{0K}^2\right]\nonumber\\
&+&\frac{24M_{0\pi}^2}{F^4}\nu_\pi(T)\left[\left(L_4^r + L_5^r - 2 L_6^r - 2 L_8^r\right)M_{0\pi}^2+2\left(L_4^r - 2 L_6^r\right)M_{0K}^2\right]\nonumber\\
&+&\frac{16M_{0K}^2}{F^4}\nu_K(T)\left[\left(L_4^r - 2 L_6^r\right)M_{0\pi}^2+\left(2 L_4^r + L_5^r - 4 L_6^r - 2 L_8^r\right)M_{0K}^2\right]\nonumber\\
&+&\frac{8}{27F^4}\nu_\eta(T)\left[\left(-3 L_4^r + L_5^r + 6 L_6^r - 48 L_7^r - 18 L_8^r\right)M_{0\pi}^4+2\left(3 L_4^r - 4 L_5^r - 6 L_6^r + 48 L_7^r + 24 L_8^r\right)M_{0\pi}^2 M_{0K}^2\right.\nonumber\\&+&\left.8\left(3 L_4^r + 2 (L_5^r - 3 (L_6^r + L_7^r + L_8^r))\right)M_{0K}^4\right]\nonumber\\
&+&\left.\frac{1}{8F^4}\left[\left(3 \hat C_1^r - 2 \hat C_2^r +\hat
      C_3^r\right)M_{0\pi}^4+4\left(\hat C_2^r - \hat C_3^r\right)M_{0\pi}^2 M_{0K}^2+4\hat C_3^rM_{0K}^4\right]
\right\},
\label{condtwosu3nnlo}
\end{eqnarray}

\begin{eqnarray}
\langle \bar s s \rangle_{NLO} (T)&=&-B_0F^2\left\{1+\frac{4}{F^2}\left[-\left(H_2^r - 4 L_6^r + 2 L_8^r\right) M_{0\pi}^2+2\left(H_2^r + 4 L_6^r + 2 L_8^r\right) M_{0K}^2  \right]-4\mu_K(T)-\frac{4}{3}\mu_\eta(T) \right\},\label{condsnlo}\\
\langle \bar s s \rangle_{NNLO} (T)&=&\langle \bar s s \rangle_{NLO}-B_0F^2\left\{
\frac{8}{9}\mu_\eta^2(T)-\frac{8}{3}\mu_K(T)\mu_\eta(T)
+\frac{1}{F^2}\left[\frac{4}{3}M_{0\pi}^2\mu_\pi(T)\nu_\eta(T)\right.\right.\nonumber\\
&-&\left.\frac{32}{9}M_{0K}^2\mu_K(T)\nu_\eta(T)-\frac{8}{3}M_{0K}^2\mu_\eta(T)\nu_K(T)
+\frac{4}{27}\left(16M_{0K}^2-7M_{0\pi}^2\right)\mu_\eta(T)\nu_\eta(T)\right]\nonumber\\
&+&\frac{48}{F^2}\mu_\pi(T)\left(L_4^r - 2 L_6^r\right)M_{0\pi}^2
\nonumber\\&+&\frac{32}{F^2}\mu_K(T)\left[\left(L_4^r - 2 L_6^r\right)M_{0\pi}^2+2\left(2 L_4^r +  L_5^r - 4 L_6^r - 2 L_8^r\right)M_{0K}^2\right]\nonumber\\
&+&\frac{16}{9F^2}\mu_\eta(T)\left[\left(3 L_4^r - 4 L_5^r - 6 L_6^r + 48 L_7^r + 24 L_8^r\right)M_{0\pi}^2+8\left(3 L_4^r + 2 (L_5^r - 3 (L_6^r + L_7^r + L_8^r))\right)M_{0K}^2\right]\nonumber\\
&+&\frac{32M_{0K}^2}{F^4}\nu_K(T)\left[\left(L_4^r - 2 L_6^r\right)M_{0\pi}^2+\left(2 L_4^r + L_5^r - 4 L_6^r - 2 L_8^r\right)M_{0K}^2\right]\nonumber\\
&+&\frac{32}{27F^4}\nu_\eta(T)\left[\left(-3 L_4^r + L_5^r + 6 L_6^r - 48 L_7^r - 18 L_8^r\right)M_{0\pi}^4+2\left(3 L_4^r - 4 L_5^r - 6 L_6^r + 48 L_7^r + 24 L_8^r\right)M_{0\pi}^2 M_{0K}^2\right.\nonumber\\&+&\left.8\left(3 L_4^r + 2 (L_5^r - 3 (L_6^r + L_7^r + L_8^r))\right)M_{0K}^4\right]\nonumber\\
&+&\left.\frac{1}{4F^4}\left[\left(\hat C_2^r - 2 \hat C_3^r + 3 \hat C_4^r\right)M_{0\pi}^4+4\left(\hat C_3^r - 3 C_4^r\right)M_{0\pi}^2 M_{0K}^2+12\hat C_4^r M_{0K}^4\right]\right\},
\label{condsnnlo}
\end{eqnarray}
where  the Gell-Mann-Okubo relation $3M_{0\eta}^2=4M_{0K}^2-M_{0\pi}^2$ for
the $SU(3)$ leading order masses has been used and the renormalized $L_i^r,l_i^r$ and $\hat c_i^r,\hat C_i^r$ constants depend on the scale $\mu$ as explained in \cite{GomezNicola:2010tb}.
We recall that $L_4$ and $L_5$
appear because of the meson wave function
and mass renormalization.
In $SU(3)$, the constant $L_7^r$ stems from the  eta mass renormalization.

Let us remark that the above expressions for the condensates are related to those obtained for finite volume
at NNLO ChPT in \cite{Bijnens:2006ve}, after identifying the thermal functions $\mu_i(T)$ and $\nu_i(T)$ used here with the finite volume functions $-\tilde A/2F^2$ and $-\tilde B/2$ used in \cite{Bijnens:2006ve}, respectively.

\section*{Acknowledgments}
Work partially supported by the Spanish Research contracts
FIS2008-01323 and FPA2011-27853-C02-02. We acknowledge the support
of the European Community-Research Infrastructure
Integrating Activity
``Study of Strongly Interacting Matter" 
(acronym HadronPhysics2, Grant Agreement
n. 227431)
under the Seventh Framework Programme of EU.

\end{document}